\documentclass[aps,prd,floats,preprint,nofootinbib]{revtex4}

\usepackage{graphicx}
\usepackage{multirow}
\usepackage{fancyvrb}
\usepackage{makeidx}
\usepackage{latexsym,amssymb,amsmath,graphicx, makeidx} 
\usepackage{epstopdf}
\def\gev{~{\mbox{GeV}}}
\def\filt{\text{filt}}
\def\sc{\text{SC}}
\def\rref{R_\text{ref}}
\def\ln{\text{ln}}

\def\ch\text{ch}
\def\nch{N_\text{ch}}

\newcommand{\be}{\begin{equation}}
\DefineVerbatimEnvironment{code}{Verbatim}{fontsize=\small}
\newcommand{\ee}{\end{equation}}
\newcommand{\bi}{\begin{itemize}}
\newcommand{\ei}{\end{itemize}}
\newcommand{\npu}{\langle N_{\text{pu}}\rangle}
\newcommand{\mfilt}{m_{\text{filt}}}
\newcommand{\es}{\varepsilon_S}
\newcommand{\eb}{\varepsilon_B}
\newcommand{\eseb}{\varepsilon_S/\sqrt{\varepsilon_B}}
\newcommand{\subtr}{\text{subtr}}

\begin{document}
\title{Jet Radiation Radius}
\author{Zhenyu Han}
\affiliation{\small \sl Institute for Theoretical Science, University of Oregon, Eugene, OR 97403, USA}

\begin{abstract}
Jet radiation patterns are indispensable for the purpose of discriminating partons' with different quantum numbers. However, they are also vulnerable to various contaminations from the underlying event, pileup, and radiation of adjacent jets. In order to maximize the discrimination power, it is essential to optimize the jet radius used when analyzing the radiation patterns. We introduce the concept of jet radiation radius which quantifies how the jet radiation is distributed around the jet axes. We study the color and momentum dependence of the jet radiation radius, and discuss two applications: quark-gluon discrimination and $W$ jet tagging. In both cases, smaller (sub)jet radii are preferred for jets with higher $p_T$'s, albeit due to different mechanisms: the running of the QCD coupling constant and the boost to a color singlet system. A shrinking cone $W$ jet tagging algorithm is proposed to achieve better discrimination than previous methods.
\end{abstract}
\maketitle
\thispagestyle{empty}

\setcounter{page}{1}
\pagenumbering{arabic}


\section{Introduction}
\label{sec:introduction}
It often happened in high energy physics that previously discovered particles later became the ``standard candles'', and led to the discoveries of new particles. For example, the $W$ and $Z$ gauge bosons were discovered by identifying the leptons they decay to; they (especially the $Z$ boson) in turn played important roles in the discovery of the Higgs boson at the Large Hadron Collider (LHC) \cite{higgs_discovery}. It is our hope that at the LHC and future colliders, we will be able to find new physics beyond the standard model (SM), again by first identifying particles that we are already familiar with. Therefore, it is crucial that we are able to efficiently identify all known particles, or, equivalently, their masses and quantum numbers. This is a difficult task when the particle is a QCD parton, namely a quark or a gluon, or when it decays to QCD partons. In these cases, what we see in a collider detector is one or more energetic jets that contain multiple hadrons, and their identities are concealed.

There are several handles one can utilize to distinguish jets from different origins: the jet mass or multi-jet invariant mass is determined by the original parton's mass (for convenience, in this article we use the term ``parton'' to refer to either a QCD parton or a massive particle that decays to QCD partons); a massive hadronically decaying particle, such as a $W$ or $Z$ boson, results in more than one jets with similar momenta, which is different from a usual QCD jet originated from a single quark or gluon; finally, the QCD quantum number determines the amount of jet radiation and how it is distributed. The first two handles have been used routinely in high energy experiments, while the last, though receiving significant amount of studies recently\footnote{See Ref.~\cite{tasi} for a review and more references.}, has not been used as commonly. Part of the reason is, compared with the hard kinematics, the radiation information of a jet is often associated with soft particles, making it more vulnerable to various contaminations, including the underlying event, pileup, as well as radiation from nearby jets. We will focus on tackling this difficulty in this article. 

Variables sensitive to the jet radiation are calculated from jet constituents. On the one hand, we would like to choose a larger jet radius such that more radiation from the initial parton is included  to give us more color information; on the other hand, we would like to avoid introducing excessive contamination, and a smaller radius is preferred. Therefore, a careful choice of the jet radius is important. Similar considerations have been made in Refs.~\cite{ua2, snowmass, Cacciari:2008gd, Soyez:2010rg}. In Refs.~\cite{ua2, snowmass}, it is shown that the optimum jet radius is around $R=0.7$ for jet $p_T\sim O(100\gev)$ at a hadron collider, where the effects of the underlying event and hadronization are minimized, and it yields the optimum $W/Z$ mass measurement for the UA2 experiment. The study has been extended to the LHC in Refs.~\cite{Cacciari:2008gd, Soyez:2010rg} where the criterion for optimizing the jet radius is to obtain the best mass resolution for new massive hadronically decaying resonances. The purpose of those studies is to properly reconstruct the kinematics, {\it i.e.}, the momenta of the partons initiating the observed jets. Our goal is different: we want to choose the jet radius such that the discrimination power, including that from the jet radiation patterns, is maximized. For this purpose, it is important to study the distribution of the ``intrinsic'' radiation, {\it i.e.}, the radiation originated from the initial parton instead of the contaminations. In particular, it is helpful to know what is the amount of the intrinsic radiation contained in a cone of a particular size around the jet axis. Nonetheless, there is not a unique definition of the ``amount'' of radiation, and the radiation distribution depends on various factors such as the color configuration, the boost to the system, {\it etc.}. 

In this article, we give a rigorous definition of the jet {\it radiation radius}, taking the simplest case as the start point: a dijet color singlet system. Roughly speaking, the jet radiation radius, $R(x)$, is the jet radius one needs to include on average a fraction of $x$ of the total radiation in the jet. A variable that quantifies the amount of radiation is needed in this definition, such as thrust, girth \cite{quark-gluon, quark-gluon-2}, or $N$-subjettiness \cite{nsubjettiness}. We study the dependence of $R(x)$ on the jet QCD quantum number and the jet momentum. We examine two applications, quark-gluon discrimination and $W$ jet tagging, and show that knowledge of the jet radiation radius can help us achieve the optimum discrimination power. Interestingly, in both cases, the jet radiation radius, and thus the optimum (sub)jet radius decreases with increasing jet momentum, albeit for different reasons. For a quark or gluon jet produced from a color singlet system in its (nearly) rest frame, the jet momentum is roughly proportional to the center of mass energy, which determines the amount of radiation and how it is distributed. It is well known that the QCD coupling constant decreases when the energy scale increases, resulting in smaller radiation radius. We will see that the radiation radius approximately follows a power law dependence on the jet momentum, with the power being around $-0.5\sim -0.2$. Therefore, to reduce the contaminations, we are able to use a smaller cone size to evaluate jet radiation variables for higher jet momentum.

For the case of a boosted $W$ boson (or other massive color singlet particles), the radiation scale is fixed to the $W$ mass, while the boost makes the radiated particles collimated. This results in a radiation radius inversely proportional to the jet momentum, or subjet momentum if the $W$ is identified as a single jet. In light of that, we propose a new $W$ tagging algorithm to deal with high pileup. In this algorithm, we start as usual with a fat jet that contains a $W$ boson or its QCD counterpart, and use a jet grooming method \cite{filtering, pruning, trimming} to reconstruct the kinematics of the $W$ decay. Then, when calculating a jet radiation variable, we use two much smaller cones around the two leading subjets' axes, with the cone sizes inversely proportional to the subjets' $p_T$'s. With this ``shrinking cone'' algorithm, we find a 30\% (60\%) improvement in the statistical significance for jet $p_T=300\gev$ ($150\gev$), when the average number of pileup events is 60, compared with methods not using the shrinking cones. 

The rest of the article is organized as follows. In Section \ref{sec:definition}, we give the definition of the jet radiation radius. In Section \ref{sec:quark-gluon}, we discuss the difference in the radiation radius between a quark jet and a gluon jet in a dijet color singlet system, and demonstrate that the optimum jet radius for quark-gluon discrimination decreases for increasing jet $p_T$. Section \ref{sec:wtag} is devoted to the discussion of the radiation radius in a boosted system, and in particular its role in $W$ jet tagging. Section \ref{sec:discussions} contains some discussions.  

\section{Definition}
\label{sec:definition}
A jet has a finite radius precisely because of the QCD radiation and hadronization. Therefore, it seems redundant to have a definition of a {\it radiation radius}. However, the usual purpose of using a finite radius for jet clustering is to  ensure most of the jet radiation is included in the jet such that one can infer the initial partons' momenta correctly. For this purpose, the radiation gives us difficulties rather than provides us useful information. Moreover, due to collinear singularity, the momentum of a jet is concentrated in the 'core', that is, the center of the jet. Therefore, it is not essential to control precisely the jet radius, and multiple choices of jet radius coexist in high energy experiments. For the purpose of determining the quantum number of a jet, the information from the QCD radiation is essential and one should be more careful about choosing the optimum jet radius. As mentioned in the introduction, this is particularly important when the jet is contaminated by other hadronic activities in the event, including the underlying event, pileup and   radiation from nearby jets. Therefore, it is useful to define a jet radiation radius, which is a measure of how the radiation is distributed. Knowing the ``intrinsic'', {\it i.e.,} uncontaminated, jet radiation radii corresponding to partons with different quantum numbers will allow us to have a generic understanding of how to choose a jet radius to better deal with the contaminations. As we will show later, it also allows us to engineer new algorithms that provide better discrimination powers.

The first jet definition was given by Sterman and Weinberg \cite{sterman-weinberg}. In Ref.~\cite{sterman-weinberg}, a hadronic event from an $e^+e^-$ collision is classified as a two jet event if at least a fraction of $1-\epsilon$ of the event's total energy is contained in two cones of opening half-angle $\delta$, where $\epsilon$ is a small fraction. Obviously, for fixed $\epsilon$, when we increase $\delta$, the fraction of hadronic events that are classified as two jets, $f_2$, will also increase. At the next leading order, $f_2$ is given by \cite{qcd-collider}
\begin{equation}
f_2=1-8C_F\frac{\alpha_S}{2\pi}\left\{\ln\frac1\delta\left[\ln\left(\frac1{2\epsilon}-1\right)-\frac34+3\epsilon\right]+\frac{\pi^2}{12}-\frac{7}{16}-\epsilon+\frac{3}{2}\epsilon^2+O(\delta^2\ln\epsilon)\right\}.
\label{eq:f2}
\end{equation}
When $\epsilon$ is small, we keep only the leading term and find 
\begin{equation}
\delta_q\sim\text{exp}\left[-\frac{\pi(1-f_2)}{4C_F\alpha_S(s)\ln(1-\epsilon)}\right].
\label{eq:deltaq}
\end{equation}
In Eq.~(\ref{eq:deltaq}), we have added a subscript  ``$q$'' because hadronic events in an $e^+e^-$ machine are dominated by quark-antiquark pairs. In such a machine, the event is not contaminated by the underlying event and initial state QCD radiation\footnote{It is still possible to have initial state electroweak radiation.}, and two-jet events are dominantly initiated from a pair of back-to-back high energy quarks. Therefore, $\delta_q$ can be viewed as the ``intrinsic'' size of a quark jet\footnote{Of course, this is incorrect for special configurations. For example, a hard gluon, containing about half of the total energy, may be occasionally emitted at a large angle from both of the two quarks and classified as one of the two jets. However, in practice, this rarely happens and does not affect our studies of the properties of the quark jet.}. From Eq.~(\ref{eq:deltaq}), we see the definition of the size $\delta_q$ depends on two parameters, $\epsilon$ and $f_2$. Similarly, if the system under study is initiated from two hard gluons (in the color singlet configuration), we obtain the angular size, $\delta_g$, of a gluon jet. At the leading order, $\delta_g\sim \delta_q^{C_F/C_A}$ \cite{qcd-collider}. Therefore, $\delta_q$ or $\delta_g$ is a measure of how the radiation is distributed, which is sensitive to the partons' color structure. This motivates us to give a general definition of a jet {\it radiation} radius, by replacing the energy fraction $1-\epsilon$ with a measure of the amount of radiation. The definition of the amount of radiation is not unique, and suitable choices include thrust (denoted $T$), charged particle multiplicity (denoted $\nch$), $N$-subjettiness, {\it etc}.

While a possible definition of the jet radiation radius could be given following the definition of the variable $\delta$, in practice it is perhaps more convenient to use the following definition.

{\it In a 2(or $N$)-jet color singlet system in its center of mass frame, a jet radiation radius is defined as the jet radius that the average amount of radiation within the 2 (or N) jets is a fraction of $x$ of the average total radiation.}

In practice, the jet radiation radius can be calculated as follows. For a given ensemble of events, to get the fraction $x$, we first calculate the average ``total'' radiation: for an $e^+e^-$ machine, it may include all particles in the event; while for a hadron collider, we may use a set of jets clustered with a large radius as the start point. Then we use a (smaller) radius $R$ to recluster the jets and find the amount of radiation within the leading 2 (or N) jets, also averaged over the ensemble. The fraction $x$ is the radio of the two averages, and the corresponding $R$ is our jet radiation radius. We denote the radiation radius by $R_{var}(x)$, where $var=1-T, \nch, \tau_1 \ldots$, is a measure of the amount of radiation.
 
A few comments of the above definition is in place. First, there are multiple choices of the variables that can quantify the amount of radiation. To be a good candidate, the variable should satisfy: $a.$ it is (near) zero when there is no radiation. $b.$ it increases monotonously when the jet radius is increased, and approaches a finite value when the jet radius is taken to be infinity, {\it i.e.}, when all particles in the event (or the part of the event being studied) are included. We easily see that among the most commonly used variables, jet mass is a good radiation variable, while jet momentum is not. Other radiation variables include $1-T$ where $T$ is the thrust, charged particle multiplicity, $N$-subjettiness, {\it etc}.
Second, the requirement of a color singlet system is to guarantee the jet radius is not affected by particles outside the system, otherwise the definition is ambiguous. Moreover, we have also required the system to be in the center of mass frame such that it is reasonable to use a common jet radius for all jets in the system. Even with this requirement, when the system contains more than 2 jets with different momenta, we still need a prescription on how to choose the jet radii -- this is not a concern in this article because we will only discuss dijet systems. As we will see, when a dijet system is boosted, the two jets will have different momenta and it is desirable to use different radii for them.

In this article, we will rely on Pythia 8 \cite{pythia8} simulations to calculate the jet radiation radius, leaving analytical calculations to future studies.

\section{Dijets: Quark and gluon discrimination}
\label{sec:quark-gluon}
We start from the simplest case: dijet color-singlet systems. There are two possibilities for the initial partons, $q\bar q$ or $gg$. Given the large number of variables that have been proposed in the literature, we will only consider a few of them that are particular useful for our later discussions: charged particle multiplicity, girth and $N$-subjettiness. Charged particle multiplicity is simply the number of charged particles passing a certain $p_T$ threshold, or the number of tracks in the jet. Girth is defined as \cite{quark-gluon}
\begin{equation}
g\equiv\sum_{i\in\text{jet}}\frac{p_T^i}{p_T^{\text{jet}}}r_i,
\end{equation}   
where $r_i$ is the distance in the $(y,\phi)$ plane between particle (or jet constituent) $i$ and the jet axis. Here, $y$ is the rapidity. Note the variable contains a normalization factor, $p_T^{\text{jet}}$, which is the jet $p_T$. As described in the previous section, we will need to calculate the ``total radiation'' starting from a fat jet (or, for the $e^+e^-$ case, starting from all particles in a hemisphere), as well as the radiation for a variety of smaller jet radius. Then it is a question whether we should use the $p_T$ of the fat jet (or half of the center of mass energy for an $e^+e^-$ machine)  as a common normalization factor for all jet radii, or use the jet $p_T$ of the corresponding radius. Both choices are valid, although there are subtle differences between them. We find the latter is more convenient and more commonly used, which we will stick to in the following discussions. 

$N$-subjettiness is derived from $N$-jettiness \cite{njettiness}, and defined as follows \cite{nsubjettiness}. We define a distance measure between a particle $k$ and an axis $J$ as $\Delta R_{J,k}^\beta$, where $\Delta R=\sqrt{\Delta\eta^2+\Delta\phi^2}$, and $\beta$ is a constant. Then for $N$-axes we define
\begin{equation}
\tilde\tau_N=\frac{1}{d_0}\sum_kp_{T,k}\min\left\{\Delta R_{1,k}^\beta,\Delta R_{2,k}^\beta,\cdots,\Delta R_{N,k}^\beta\right\}.
\end{equation}
Namely, for each particle in the jet, we find the nearest axis and calculate the $p_T$ weighted distance. Then $\tilde\tau_N$ is a sum of the distances over all particles in the jet, normalized by a factor 
\begin{equation}
d_0=\sum_ip_{T,i}(R_0)^\beta,
\end{equation}
where $R_0$ is the jet radius. Finally, we find the $N$-axes such that $\tilde\tau_N$ is minimized, and the minimum value is called $N$-subjettiness and denoted $\tau_N$. 

When $N=1$ and $\beta=1$, we see that 1-subjettiness, $\tau_1$, is defined very similar to girth, with slight differences in the distance measure (girth uses the rapidity $y$ while $\tau_1$ uses the pseudorapidity $\eta$) and the normalization factor. Therefore, the discussion on girth applies to $\tau_1$ as well. In general, these variables are examples of a set of variables called {\it radial moments} \cite{quark-gluon-2} that are all valid radiation variables:
\begin{equation}
M_f=\sum_{i\in \text{jet}}\frac{p_T^i}{p_T^\text{jet}}f(r_i),
\end{equation}
where $f(r)$ is a function of $r$, such as $r$, $r^2$, $r^3$\ldots. Although it will be infrared/collinear unsafe, we may also generalize the definition to allow a non-linear dependence on $p_T^i$, for example, by making the sum over $p_{T,i}^\alpha r_i^\beta$. Then the particle multiplicity can be viewed as a special case where $\alpha=\beta=0$. 

\begin{figure}[hbt!]
\centering
\begin{tabular}{cc}
\includegraphics[width=0.5\textwidth]{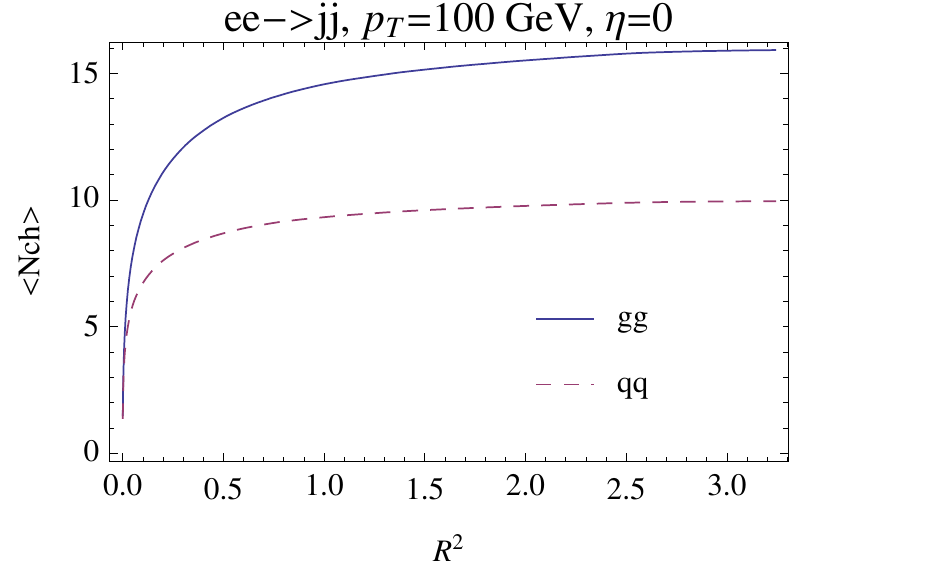}&
\includegraphics[width=0.5\textwidth]{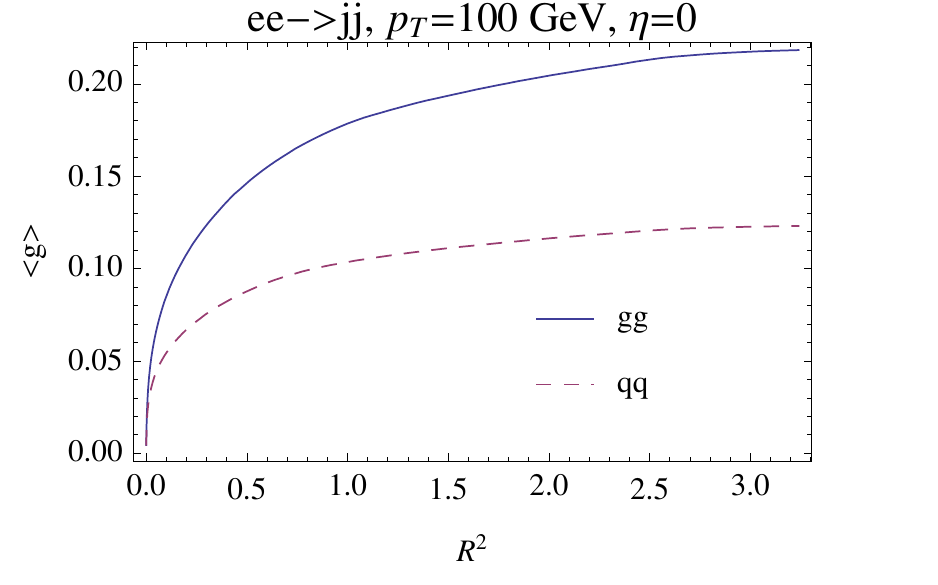}
\end{tabular}
\caption{The average values of charged particle multiplicity ($\nch$) and girth ($g$) as a function of $R^2$.\label{fig:vars_ee} }
\end{figure}

As mentioned in the previous section, it is convenient to study events from $e^+e^-$ collisions, where we only have final state QCD radiation emitted from the partons being studied. Therefore, we study the processes $e^+e^-\rightarrow q\bar q$ and $e^+e^-\rightarrow gg$. We fix the pseudorapidity of the two outgoing partons to $\eta=0$, {\it i.e.}, perpendicular to the beam, and use Pythia 8 for showering and hadronization. For an $e^+e^-$ machine, it is better to use the spherical coordinates. Nevertheless, we eventually would like to extend our results to a hadron collider, so we will use the $(\eta, \phi)$ coordinates for convenience. We then apply the anti-$k_t$ algorithm with various $R$'s for jet clustering, using FastJet \cite{fastjet}. It is known that the anti-$k_t$ algorithm gives us circular jets, which mimic cones of size $R$ around the jet axes. Since all contaminations are proportional to the area of the jet, we draw the radiation variables as a function of $R^2$ in Fig.~\ref{fig:vars_ee} for better illustration. It is clear from Fig.~\ref{fig:vars_ee} that the radiation variables grow much faster at smaller $R^2$'s than larger $R^2$'s, and eventually saturate at very large $R^2$'s. This indicates that increasing $R$ beyond a certain value will only introduce more contaminations rather than provide more information, therefore, an optimum radius should exist. 

\begin{figure}[hbt!]
\centering
\begin{tabular}{cc}
\includegraphics[width=0.5\textwidth]{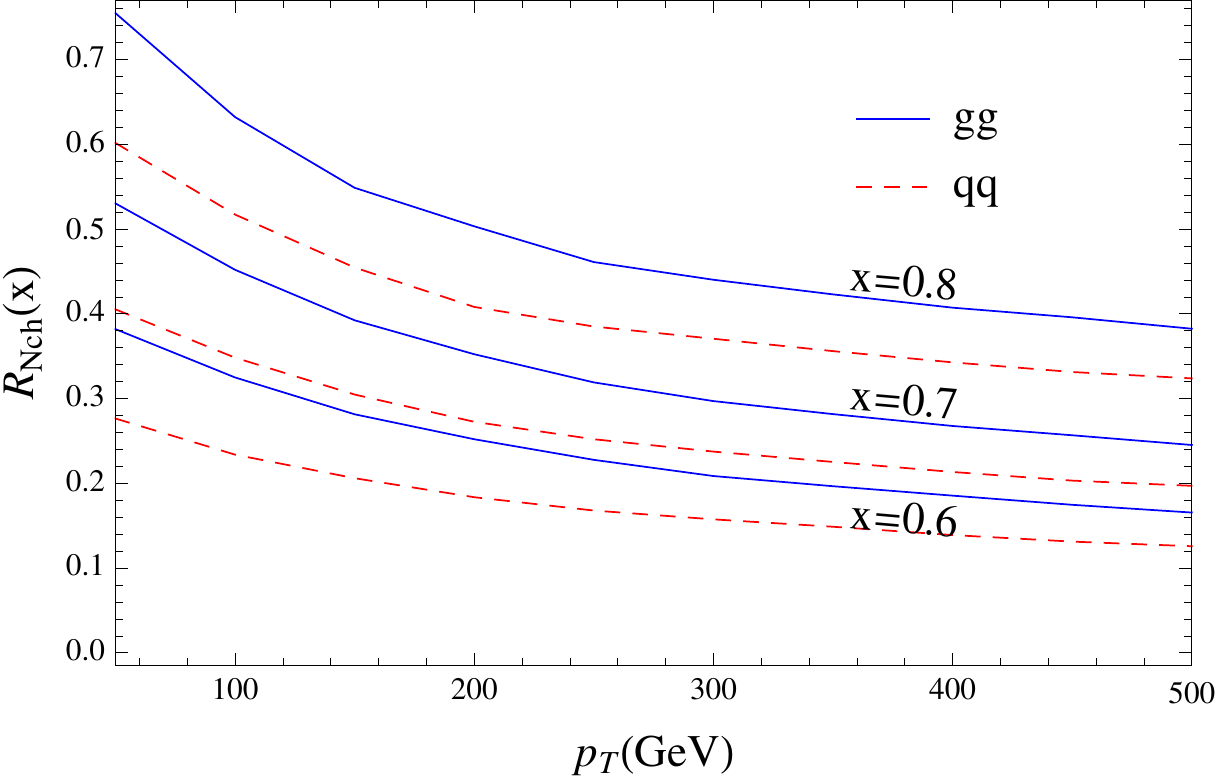}
&\includegraphics[width=0.5\textwidth]{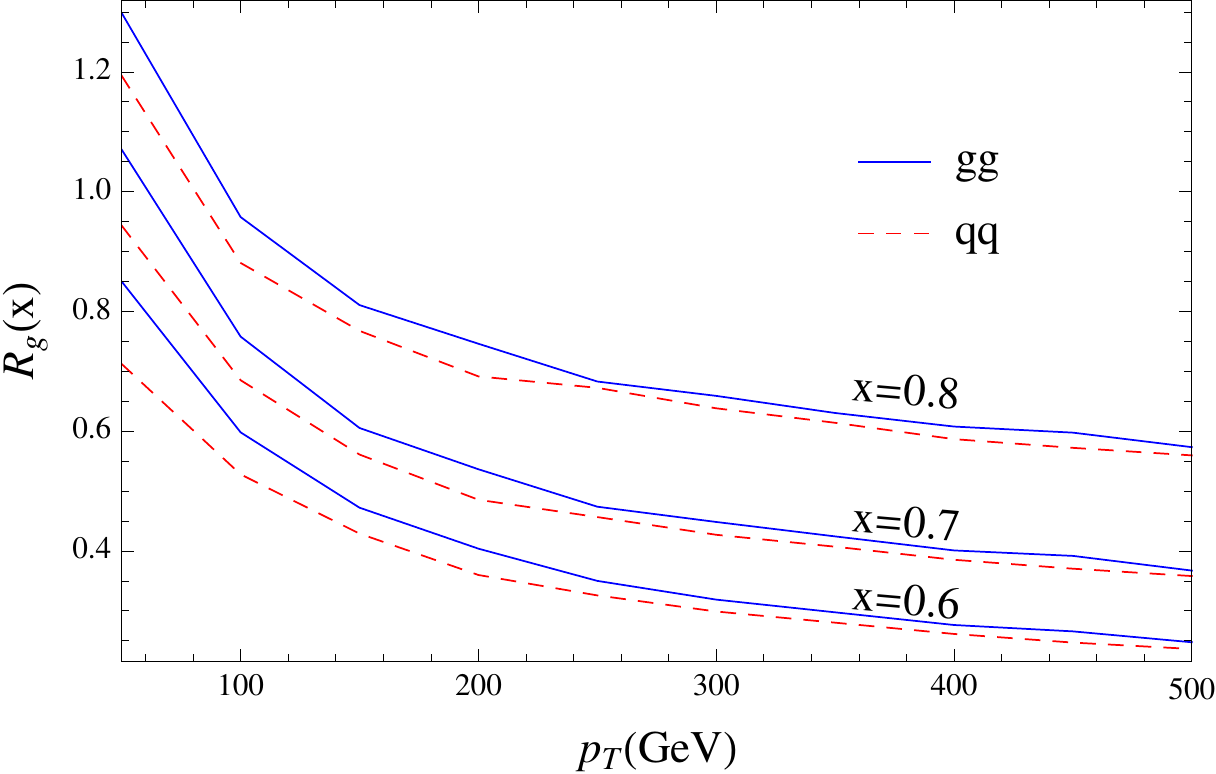}\\
\end{tabular}
\caption{The jet radiation radius $R_{\nch}(x)$ and $R_{g}(x)$ as a function of jet $p_T$, for x=0.6, 0.7, 0.8. \label{fig:rx_pt}}
\end{figure}

Given Fig.~\ref{fig:vars_ee}, it is easy to read the radiation radius $R(x)$ by locating the radius $R$ where $\langle \nch\rangle$ or $\langle g\rangle$ is $x$ times the saturation value. We then turn to the study of the $p_T$ dependence of the jet radiation radius. It is well known that the jet angular size $\delta$ decreases with the jet momentum because of the RGE running of the strong coupling constant $g_S$. This can be easily seen from Eq.~(\ref{eq:deltaq}): the exponentiation combined with the logarithmic running of $\alpha_S$ results in a power law decrease in $p_T$ for $\delta_q$. Similarly, the jet radius as defined in Section~\ref{sec:definition} also scales with the jet momentum. In Fig.~\ref{fig:rx_pt}, we show the $p_T$ dependence of the jet radiation radius for $var=\nch$ and $var = girth$. 

\begin{figure}[hbt!]
\centering
\begin{tabular}{cc}
\includegraphics[width=0.5\textwidth]{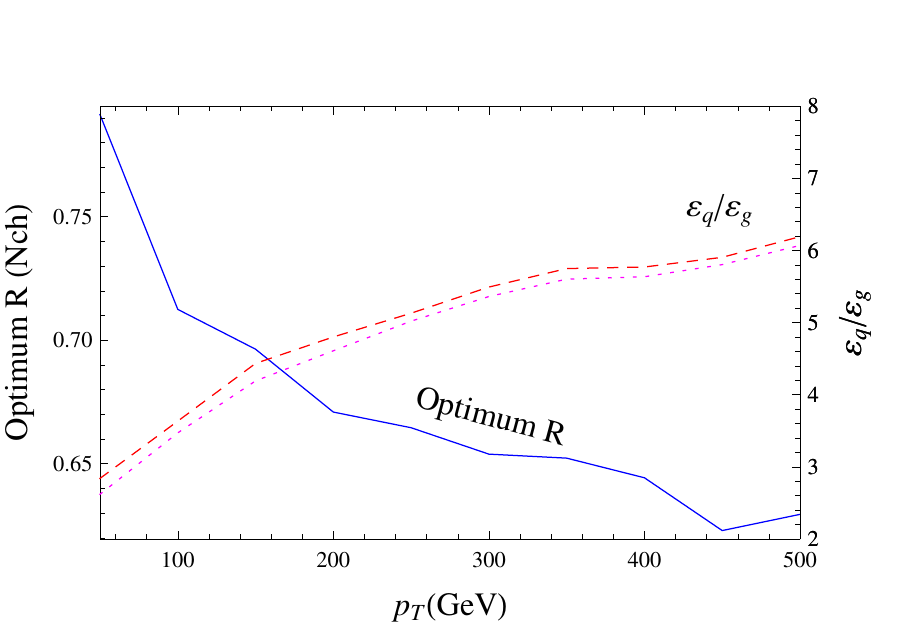}
&\includegraphics[width=0.5\textwidth]{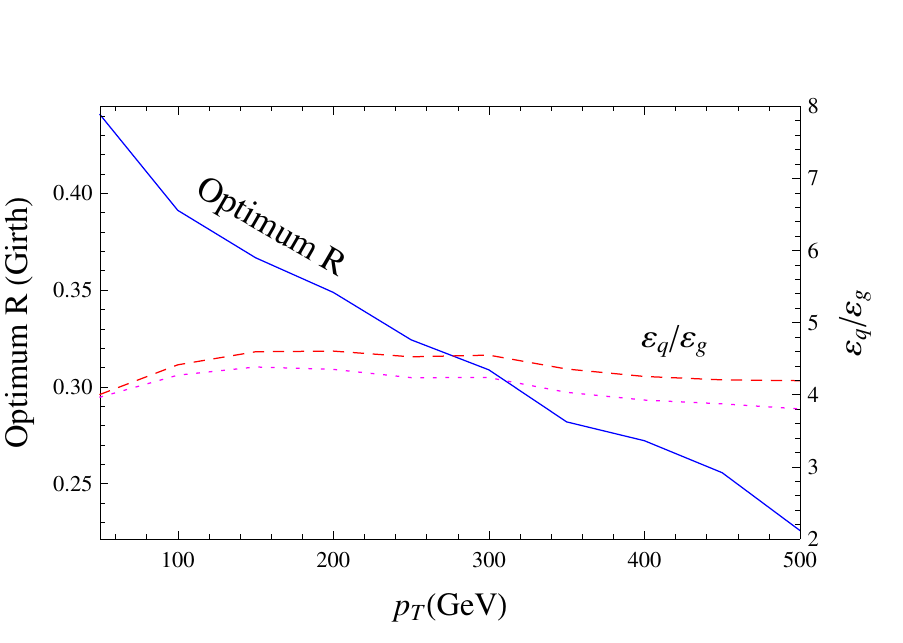}\\
\end{tabular}
\caption{The optimum jet radius for quark-gluon discrimination at the LHC (blue solid) and the corresponding efficiency ratio, $\varepsilon_q/\varepsilon_g$ (red dashed). For comparison, we also draw the efficiency ratio for fixed radius, $R=0.5$ (pink dashed). The quark jet tagging efficiency is fixed to 50\% for all cases. Left: $N_{ch}$; right: girth. \label{fig:bestR_pt}}
\end{figure}

We see that the radiation radius always decreases as the jet $p_T$ increases. The dependence on $p_T$ abides by an approximate power law with the power varying between -0.5 and -0.2 for the cases in Fig.~\ref{fig:rx_pt}. The immediate conclusion is, in the presence of contamination, if we would like to obtain the best performance for quark-gluon discrimination, we should use a shrinking radius for increasing $p_T$. At hadron colliders, this is true even without the contamination from pileup. To see this, we consider dijet events at the LHC\footnote{We refer readers to Refs.~\cite{quark-gluon, quark-gluon-2} for a comprehensive study of quark-gluon discrimination. In this article, we focus on studying how the discrimination power depends on the jet size.}.  We use Pythia 8 to generate $pp\rightarrow qq$ and $pp\rightarrow gg$ events, with the underlying event and initial/final state radiation turned on. Pileup is not included. For each choice of the dijet $p_T$, we cluster the events with varying $R$ using the anti-$k_t$ algorithm and calculate the corresponding $\nch$ and girth for the two leading jets. Only tracks with $p_T>0.5\gev$ are counted in $\nch$. The two variables are used as two separate discriminators. From Fig.~\ref{fig:vars_ee}, we see a gluon jet usually have more radiation than a quark jet, therefore, we can apply an upper cut on $\nch$ or girth to keep more quark jets than gluon jets. We fix the efficiency of the quark jets to 50\%, and evaluate the corresponding $\varepsilon_q/\varepsilon_g$, where $\varepsilon_q$ ($\varepsilon_g$) is the efficiency of quark (gluon) jets. We vary the radius between 0.1 and 1.2 and find the one that maximizes $\varepsilon_q/\varepsilon_g$ for jet $p_T$ between 50 $\gev$ and 500 $\gev$, which is plotted in Fig.~\ref{fig:bestR_pt}. 

As expected, we see in Fig.~\ref{fig:bestR_pt} that the optimum jet radius decreases for increasing $p_T$. In Fig.~\ref{fig:bestR_pt}, we have also given $\varepsilon_q/\varepsilon_g$ for a fixed jet radius, $R=0.5$, which is not significantly lower than the optimum values. Nevertheless, we do learn that a smaller than usual ($< 0.4$) jet radius is preferred for a large range of jet $p_T$ when using girth as the discriminator. This may turn out to be essential once the contamination is larger than what we have assumed, for example, when pileup is included or when we are considering a busier event topology. Moreover, we also see the optimum radii are quite different for $\nch$ and girth, which indicates if we would like to combine the two variables to maximize the distinguishing power, we should use two (or more) jet radii.  

In the above discussion, we have included all charged particles (with a $p_T$ threshold) in the calculation of $\nch$, and all charged and neutral particles when calculating girth. In a collider detector, the momenta of the charged particles can be measured with the tracking system, while those of the neutral particles have to be measured with the calorimeters. The former is less sensitive to pileup and also has a better resolution than the latter, and the two pieces of information are complementary. We have seen that even without pileup, we obtain a better discrimination power by choosing the optimum jet radius. As we will show, in the presence of pileup, a proper consideration of the radiation radius is more important, especially when we are using the information from the neutral constituents of the jet. In order to see the effects of pileup, instead of expanding upon the discussion on quark-gluon discrimination, in the next section, we consider another important application -- $W$ jet tagging in a high pileup environment. 

\section{$W$ jet tagging}
\label{sec:wtag}
\subsection{Jet radiation radius in a boosted system}
In order to avoid ambiguities, we defined the jet radiation radius for a dijet system in its center of mass frame, such that it is natural to use a common radius for the two jets. When the system is boosted, we need a prescription to decide how to choose the jet radii for jets with different momenta. As we discussed, the radiation is mostly concentrated around the jet axis. Therefore, we see that the jet radiation radius roughly scales as $1/|\mathbf{p}|$ from the following argument.

 Consider two lightlike 4-vectors, $p_1$ and $p_2$ close to the jet axis, and therefore close to each other. They are boosted to $p_1'$ and $p_2'$ under a common boost, and their opening angle $\theta$ ($\ll1$) becomes $\theta'$ ($\ll1$). Unless the boost is well aligned with the opposite direction of the jet momentum, it changes the magnitudes of the two momenta by roughly the same fraction, {\it i.e.}, 
\begin{equation*}
|\mathbf{p}_1'|/|\mathbf{p}_1|\approx|\mathbf{p}_2'|/|\mathbf{p}_2|.
\end{equation*}
Then we have 
\begin{eqnarray*}
&&p_1\cdot p_2=p_1'\cdot p_2'\\
&\Rightarrow& |\mathbf{p}_1||\mathbf{p}_2|(1-\cos\theta)=|\mathbf{p}_1'||\mathbf{p}_2'|(1-\cos\theta')\\
&\Rightarrow& \theta/\theta'\approx|\mathbf{p}_1'|/|\mathbf{p}_1|.
\end{eqnarray*}
Therefore, the angle between the two lightlike vectors is inversely proportional to their momenta. We extend this observation to a cone with multiple lightlike particles and draw the conclusion that the cone size shrinks when the jet is boosted. Of course, the inverse proportionality is only approximate, especially for radiation at large angles and for soft hadrons whose masses cannot be ignored. However, as we will show, this does not prevent us from applying the scaling relation in a realistic situation such as the $W$ tagging, and engineer new algorithms to achieve improvement.

\begin{figure}[hbt]
\centering
\includegraphics[width=0.6\textwidth]{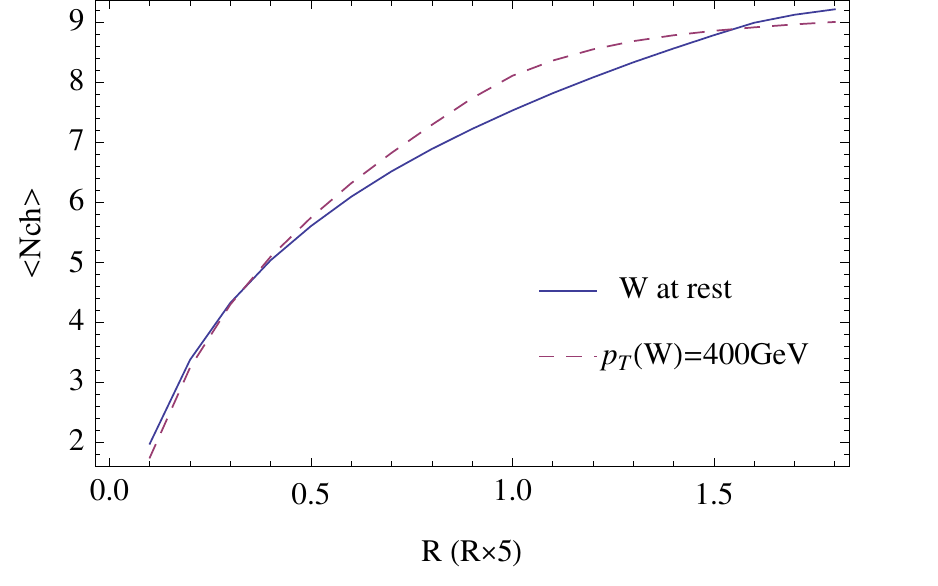}
\caption{The average number of tracks as a function of the (sub)jet radius for a $W$ at rest and a boosted $W$. The $x$ axis is the jet radius for a $W$ at rest, and 5 times the subjet radius for a boosted $W$. We set the two subjets' momenta to be equal in magnitude such that we can use the same subjet radius. The number shown is summed over the two leading (sub)jets.}
\label{fig:wdecay}
\end{figure}

For illustration, we consider the  case of a hadronically decaying $W$. Again, we first consider an $e^+e^-$ machine to avoid contaminations. We fix the kinematics of $e^+e^-\rightarrow W^+W^-$ such that the $W$'s are produced at $\eta=0$, each of which subsequently decays to two quarks with equal energy, also at $\eta=0$. Choosing the $W$'s momentum to be $400\gev$, we have the two partons from the $W$ decay each carrying about $200\gev$ momentum. Note that due to the large boost, the $W$ decay products are collimated and often clustered as a single jet. In this case, the two jets from the $W$ decay become two subjets. We then repeat the showering and hadronization and jet clustering procedure as we did for $e^+e^-\rightarrow$ dijet in Section ~\ref{sec:quark-gluon}. Comparing with a $W$ decaying in its center of mass frame, we should be able to obtain the same amount of radiation in the leading two jets for the boosted case, by using a (sub)jet radius that is 1/5 of the center-of-mass case. In Fig.~\ref{fig:wdecay}, we plot $\langle\nch\rangle$ as a function of the (sub)jet radius for the two cases, which to a good approximation confirms our expectation. Thus, we can extend our definition of the jet radiation radius to a boosted color singlet system and conclude that the radiation radius is roughly inversely proportional to the boost.

\begin{figure}[hbt]
\centering
\includegraphics[width=0.6\textwidth]{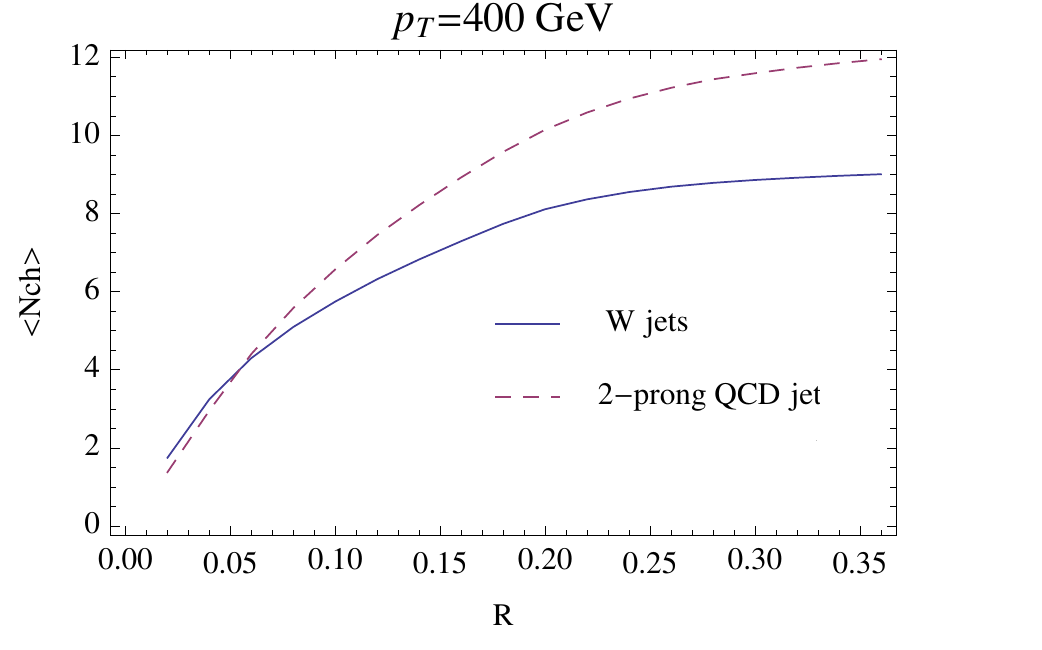}
\caption{The average number of tracks for a boosted $W$ and a 2-prong QCD jet as a function of the subjet radius, for fixed kinematic configurations (see text). The number shown is the sum of the two leading subjets\label{fig:nch_w_qcd}}.
\end{figure}

In $W$ jet tagging, after a jet grooming algorithm, the remaining background will be QCD jets that kinematically resemble a boosted $W$ decay. Besides the case that two unrelated jets accidentally merge to one fat jet, most of the background jets come from relatively hard QCD splittings. For these QCD jets, the remaining handle is the difference in the radiation patterns. We have just shown that for a boosted $W$, a smaller jet radius is needed to include most of the radiation in the jet. Then we will need to know if it is the case for a QCD jet with a similar kinematic configuration. For this purpose, we consider the process $e^+e^-\rightarrow \bar qqg$, with $q$ and $g$ having exactly the same kinematic configuration as the boosted $W$ decay discussed above. Therefore, if we use a large $R$ to cluster the event, the $qg$ pair will be clustered as a single fat jet with two hard subjets. We will call this jet a 2-prong QCD jet. Again, we shower and hadronize the event, and use various smaller $R$'s to cluster the events. We count the number of charged particles in the two subjets that correspond to the $qg$ pair, which is compared to the $W$ jet (Fig.~\ref{fig:nch_w_qcd}). From Fig.~\ref{fig:nch_w_qcd},  we clearly see that a 2-prong QCD jet tends to have more radiation than a $W$ jet. Also, we see the number of tracks in the QCD jet does not, but nearly saturates at $R=0.35$, where the $W$ jet is almost saturated. It is conceivable that we do not gain more discrimination power by going to larger $R$'s, because of the increasing contaminations. While it is as expected that the radiation is mostly contained in small cones of $R\sim 0.3$ for a boosted $W$, it is to some extent counter-intuitive for a 2-prong QCD jet -- we see in Fig.~\ref{fig:vars_ee} that $\nch$ does not saturate for a generic QCD jet even at $R\sim1.0$. Qualitatively, this can be understood using the dipole language: the QCD splitting $q\rightarrow qg$ creates a color singlet dipole. Since we have fixed the kinematics to be the same as a boosted $W$, this dipole also has a small energy scale and it behaves very similar to a boosted $W$. Therefore, the radiation from this dipole is also confined in small cones. Another color dipole exists from the initial $e^+e^-\rightarrow \bar q q$ production, which connects the 2-prong QCD jet to the jet in the opposite hemisphere. The radiation from this dipole is not confined, but only contributes to a small fraction of the radiation of the 2-prong jet. Thus, we find most of the radiation is contained in two small cones. Of course, this explanation is crude and theoretically it is very interesting to study the jet radiation distribution for special kinematic configurations.

In the above example, we have taken the two partons from the boosted $W$ decay to have the same momenta, which allows us to use the same cone size for jet clustering. It is of course not the case for a generic $W$ decay, for which we should use two different cone sizes for the two subjets, inversely proportional to their $p_T$'s. As we will show, this motivates us to design a new, improved $W$ tagging method, using different cone sizes when evaluating the radiation variables. 

\subsection{$W$ tagging with pileup}
Besides the fact that $W$ has a fixed mass, a boosted $W$ differs from a QCD (quark or gluon) jet in two other aspects. First, a $W$ jet contains two hard subjets with balanced momenta, while a QCD jet more often has only one hard subjet. Second, the $W$ boson is a color singlet particle which has a different radiation pattern from a QCD jet. A jet grooming method \cite{filtering, pruning, trimming} is efficient for exploiting the first difference, while after grooming we can further study variables sensitive to the radiation patterns. In Refs.~\cite{wtag, tracking}, we showed that the two pieces of information should be combined to obtain the optimum discrimination power. In those studies, the contamination from initial state radiation and the underlying event is included and they do not significantly affect the discrimination power. On the other hand, pileup, by which we mean multiple collisions in a beam crossing, may become the main obstacle to $W$ tagging. In particular, it has significant impact on the efficiency of the radiation variables. We illustrate this by considering two variables: the jet mass after the filtering/mass drop (MD) procedure, $m_\filt$, defined in Ref.~\cite{filtering}, and the $N$-subjettiness ratio, $\tau_{21}\equiv\tau_2/\tau_1$ \cite{nsubjettiness}. 

As shown in Ref.~\cite{tracking}, after the filtering/MD procedure, the variable $\tau_{21}$ becomes an efficient variable for measuring the amount of radiation, and it has small correlation with $m_\filt$. Therefore, we adopt a two step cut-and-count method for $W$-tagging, cutting on $m_\filt$ first and then on $\tau_{21}$. The signal ($W$-jets) efficiencies and background (QCD jets) mistag rates in the two steps are denoted $\es(m_\filt), \eb(m_\filt)$ and $\es(\tau_{21}), \eb(\tau_{21})$. The final efficiencies are the products of those in the two steps, for example, $\es(\text{final})= \es(m_\filt)\cdot\es(\tau_{21})$. Given the efficiencies, we can quantify the change in the significance by $\eseb$, {\it i.e.}, we achieve an improvement when  $\eseb > 1$.

We use Pythia 8 to simulate all-hadronic $WW$'s as our signal events and QCD dijets as the background. To simulate pileup, we turn on all soft QCD processes in Pythia 8, and add them on top of each signal/background event. The number of pileup events follows a Poisson distribution with an expectation value of $\npu$. We then find jets using FastJet \cite{fastjet} with the anti-$k_t$ algorithm ($R=1.0$). The two leading jets in each event are included in the following analysis.  

\begin{figure}[hbt!]
\centering
\begin{tabular}{cc}
\includegraphics[width=0.5\textwidth]{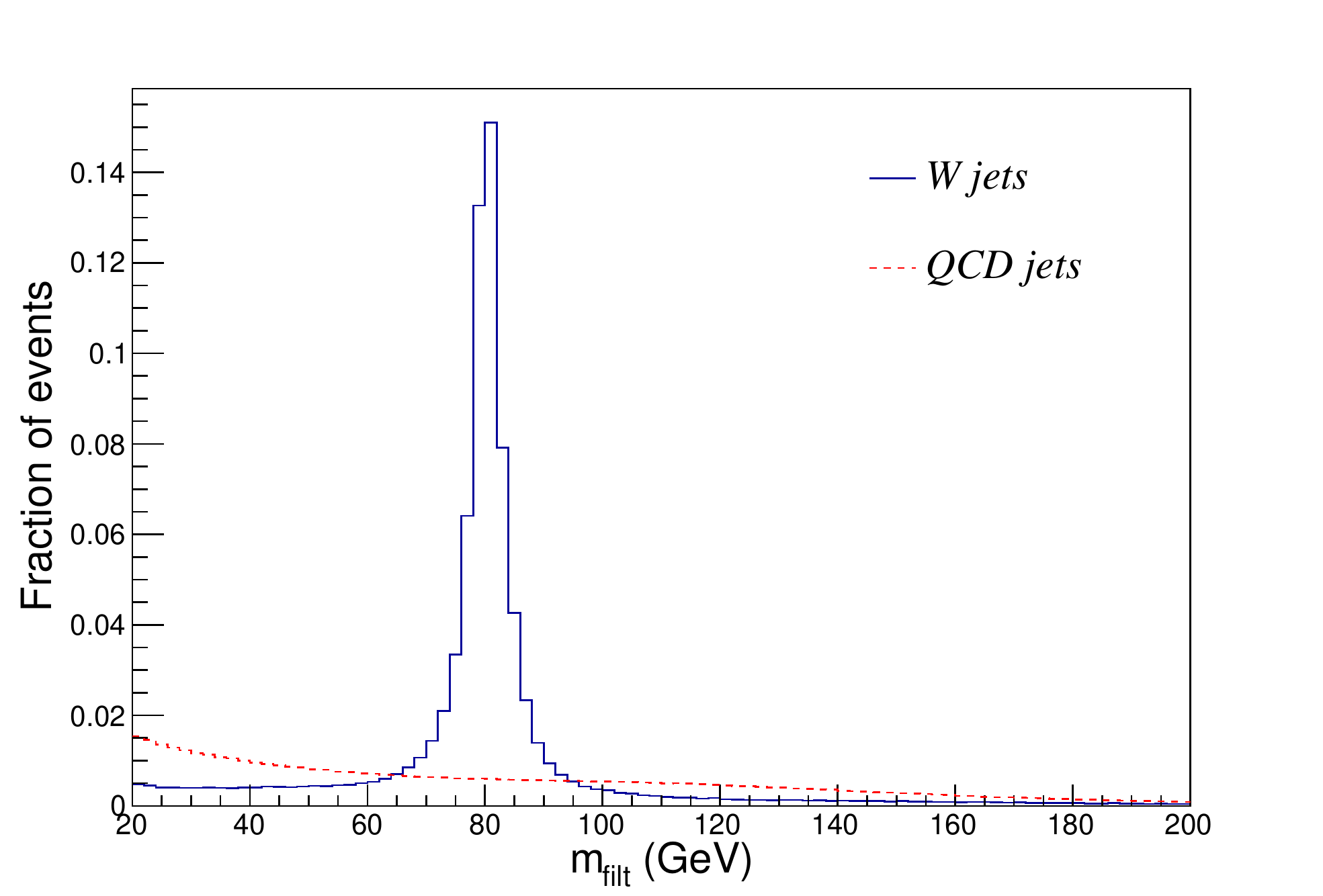}
&\includegraphics[width=0.5\textwidth]{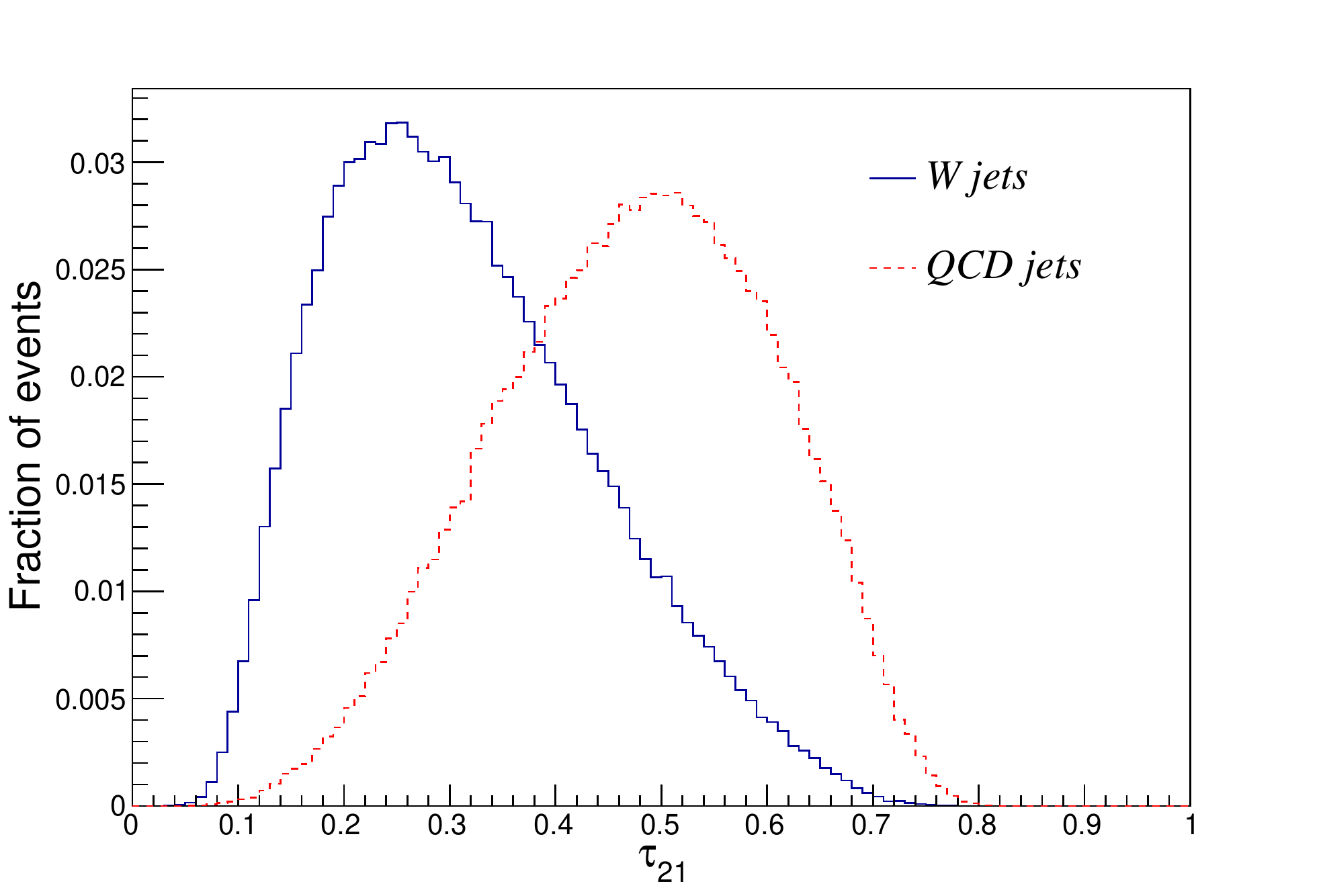}\\
\includegraphics[width=0.5\textwidth]{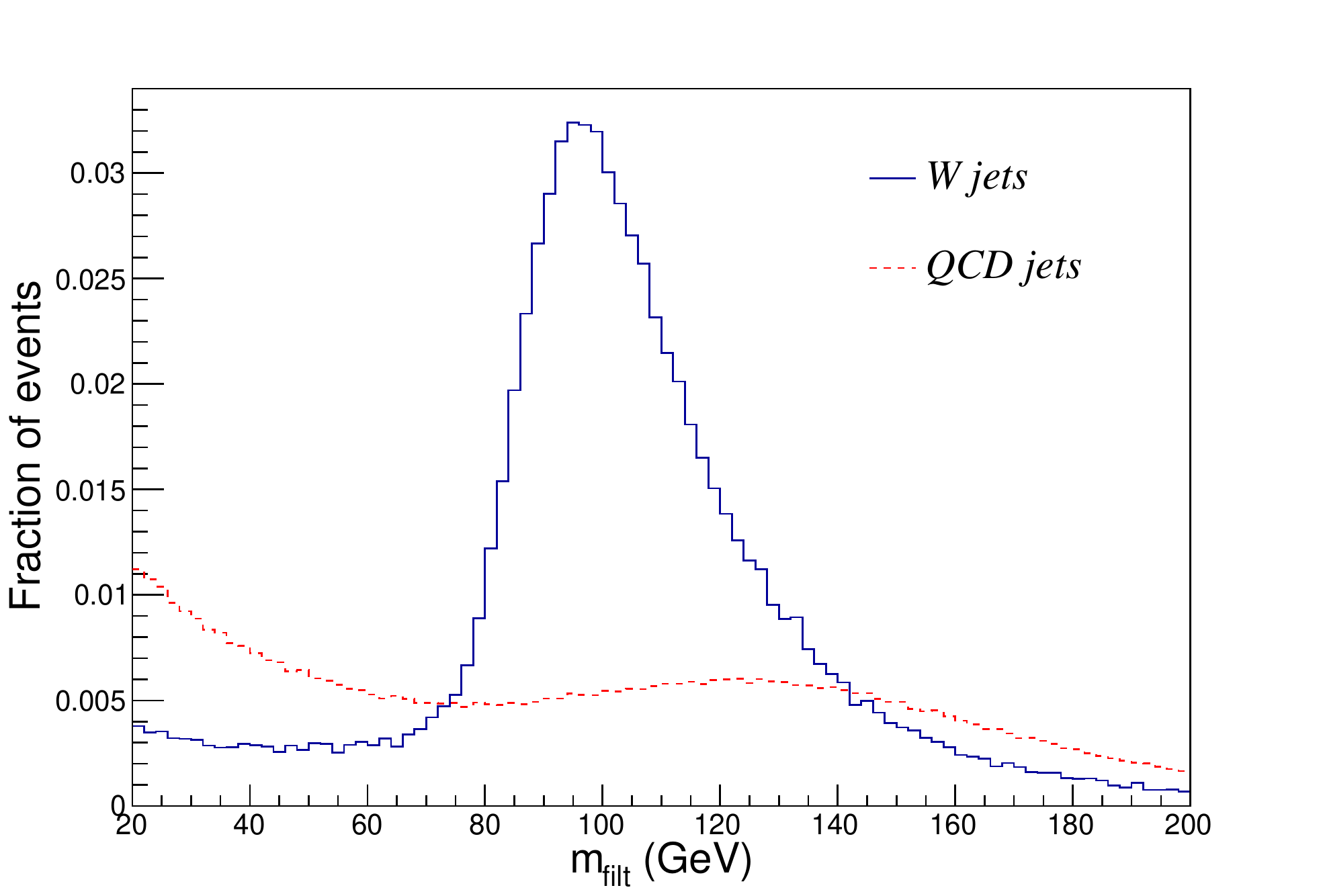}
&\includegraphics[width=0.5\textwidth]{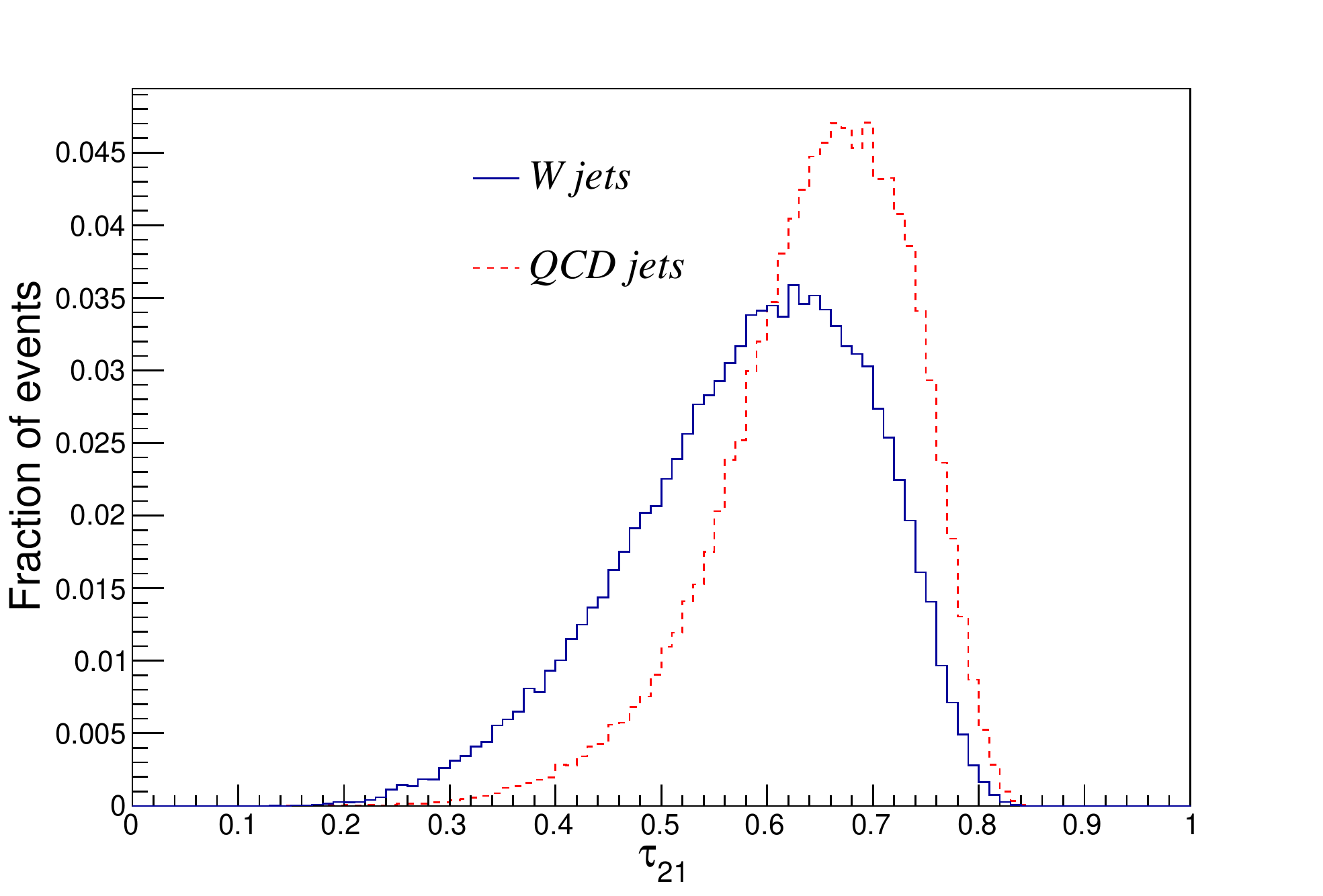}
\end{tabular}
\caption{The variables $m_{\text{filt}}$ and $\tau_{21}$ before (top) and after (bottom) including the pileup events.\label{fig:pileup}}
\end{figure}

In Fig.~\ref{fig:pileup}, we show the effect of pileup events when $\npu= 60$, for jet $p_T=300\gev$\footnote{We apply a cut $p_T>300\gev$ at parton level in Pythia 8. Due to PDF suppressions, the events are dominated by jets with $p_T$ close to $300\gev$. We take the leading two jets after jet clustering and do not further apply any $p_T$ cut at the jet level.}. We see that without pileup, the filtering/MD method is efficient to reconstruct the $W$ mass. We apply a mass window cut $(60,100)\gev$ \footnote{This is not the best mass window based on Fig.~\ref{fig:pileup}. We have used a relatively wider mass window to take into account possible experimental smearing to the reconstructed mass.} on $m_{\filt}$ and obtain a gain in the significance, $\es(m_\filt)/\sqrt{\eb(m_\filt)}=1.78$ (see Table~\ref{tab:significance}). The $\tau_{21}$ distribution after the $m_\filt$ cut is shown in Fig.~\ref{fig:pileup}. Since all jets passing the filtering/MD procedure are required to have two hard subjets, we see Fig.~\ref{fig:pileup} confirms our observation in the previous subsection that a 2-prong QCD jet tends to have more radiation than a $W$ jet. We then impose an upper cut on $\tau_{21}$, and obtain $\varepsilon_B(\tau_{21})=0.10$ at $\varepsilon_S(\tau_{21})=0.5$ -- an improvement in the significance of 1.58. This number is very close to the maximum $\es(\tau_{21})/\sqrt{\eb(\tau_{21})}$ we can get, 1.61, by scanning the $\tau_{21}$ cut, which occurs at $\es(\tau_{21})=0.38$ and $\eb(\tau_{21})=0.055$. 

After including the pileup, we see the filtered mass is shifted to larger values and the mass peak is broader. Choosing the mass window to be $(80, 120)$\gev, we obtain a signal efficiency of 0.48, and an increase in $S/\sqrt{B}$ of 1.47 which is smaller than the case without pileup. Moreover, we obtain almost no improvement by further using $\tau_{21}$: by scanning the cut on $\tau_{21}$, the biggest significance improvement is 1.02 for $\es(\tau_{21})=0.89$ and $\eb(\tau_{21})=0.76$. 

\begin{figure}[hbt!]
\centering
\begin{tabular}{cc}
\includegraphics[width=0.5\textwidth]{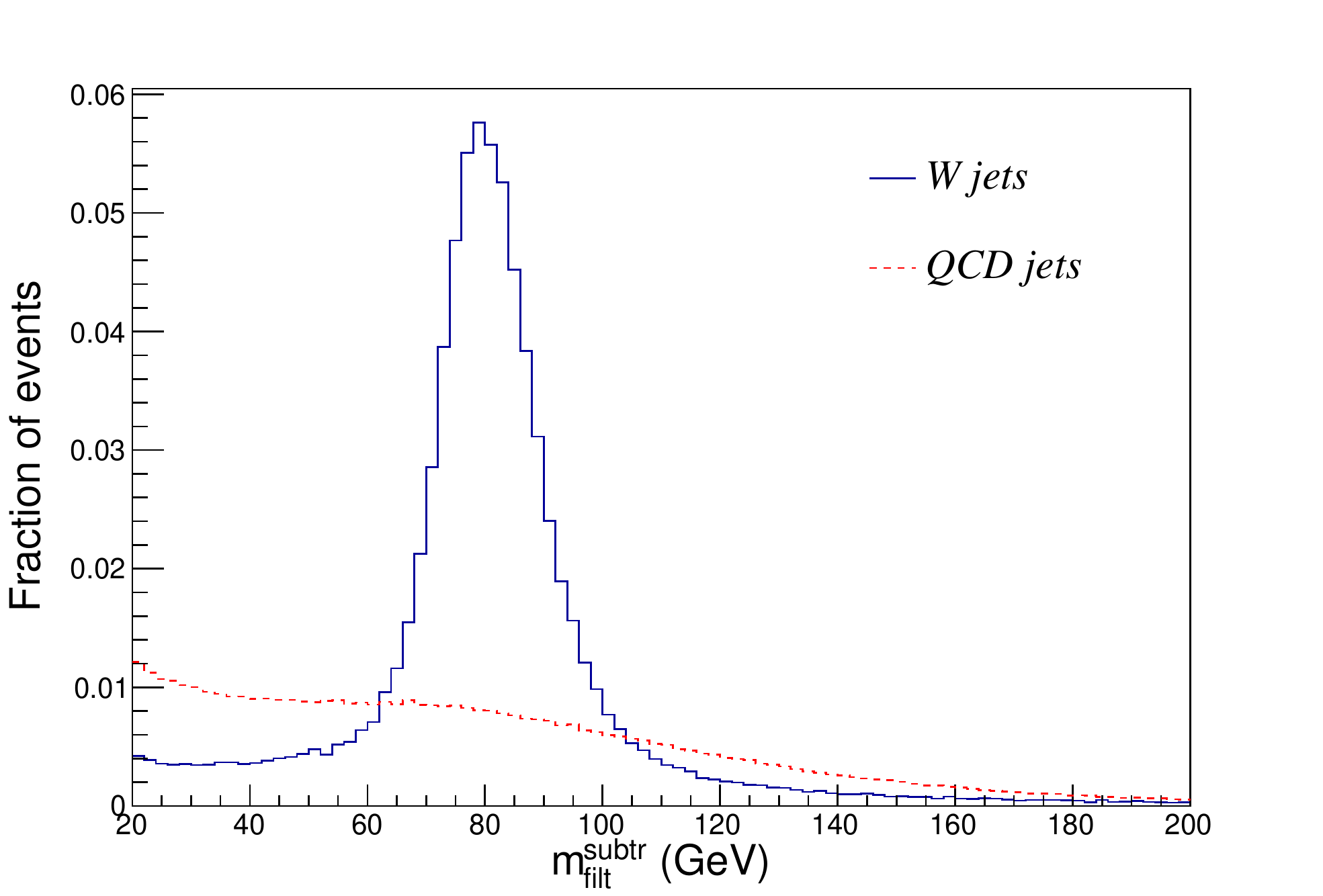}
&\includegraphics[width=0.5\textwidth]{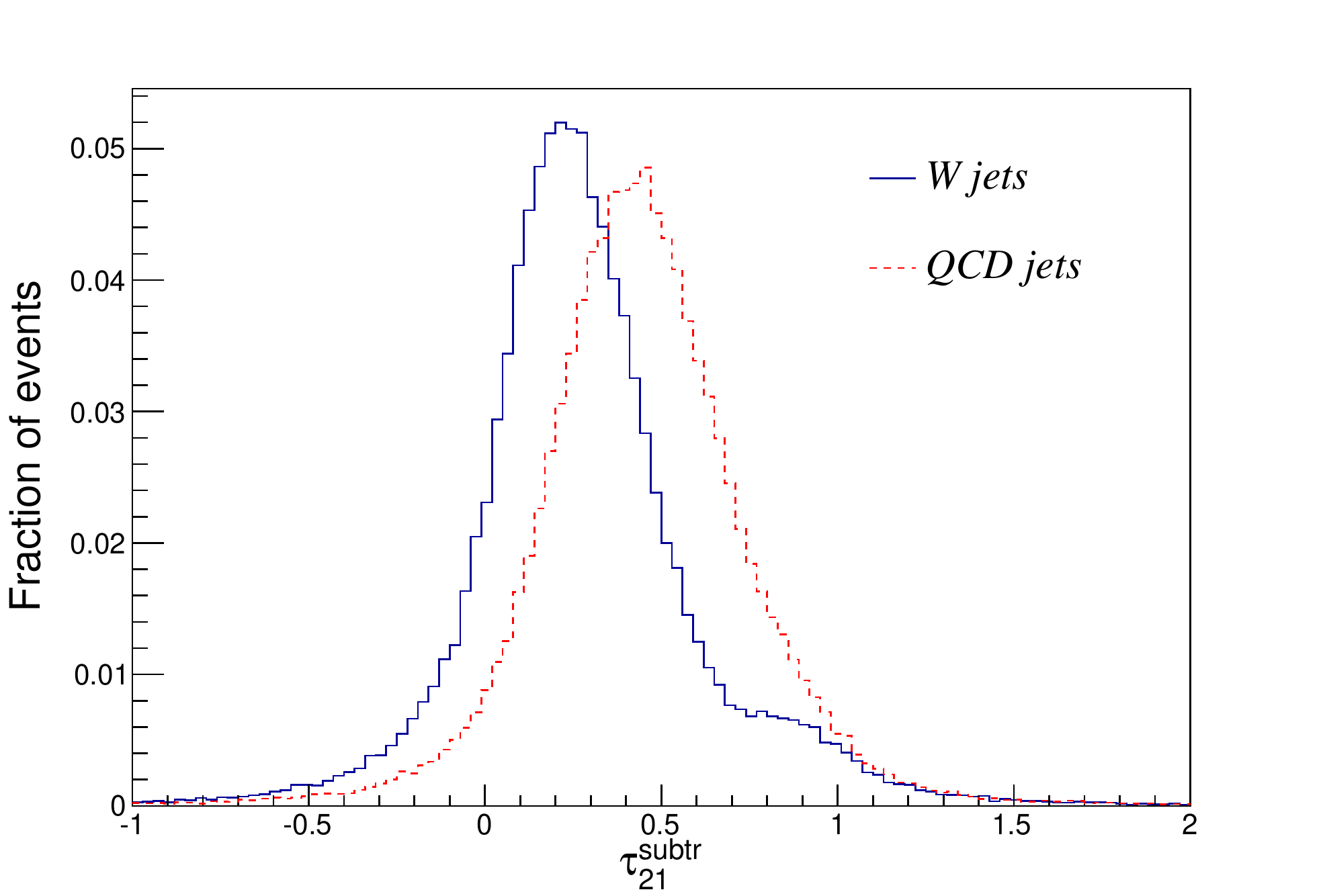}
\end{tabular}
\caption{The variables $m_{\text{filt}}$ and $\tau_{21}$ after the pileup subtraction method described in Ref.~\cite{subtraction}.\label{fig:subtraction}}
\end{figure}

In Ref.~\cite{subtraction}, a method for pileup subtraction is proposed. In this method, one first obtains the pileup $p_T$ and mass densities for a given event by dividing the event to patches and taking the medians. Then for a generic infrared and collinear safe jet shape variable, one finds its sensitivity to pileup and extrapolate to its zero pileup value using the obtained densities. Applying the method on $m_\filt$ for the $W$ jets and the QCD dijets, we find the mass peak is largely restored to its original position (Fig.~\ref{fig:subtraction}). Using the mass window $(60, 100)\gev$, we increase the significance by a factor of 1.51. On the other hand, applying the same subtraction to $\tau_{21}$, we see limited improvement. The best $\es(\tau_{21}^\subtr)/\sqrt{\eb(\tau_{21}^\subtr)}$ we can get is 1.08. This manifests that, compared with the kinematics, the radiation information is much more vulnerable to soft contaminations. Therefore, a careful consideration of the jet radiation radius is essential, which is the subject of the next subsection. 

\subsection{The shrinking cone algorithm for $W$ jet tagging}
In Section \ref{sec:quark-gluon}, we found that to optimize quark-gluon discrimination, we should use a smaller radius for a larger $p_T$. There, the reason for using a shrinking radius is the decrease of $\alpha_S$ for increasing momentum, and the dependence on $p_T$ abides by a soft power law. A boosted $W$ is similar except that the radiation radius is $\propto p_T^{-1}$, {\it i.e.}, a much stronger dependence on $p_T$. Moreover, $W$ tagging is slightly more complicated because two partons with usually different momenta are produced from the $W$ decay: first, we need to use two different radii for the two (sub)jets from the $W$ decay. Second, in practice, in order to cluster as many as $W$'s to a single jet, we often start with a large $R$ and later apply the filtering/MD procedure to identify the two subjets. This may be unnecessary because one can always use a smaller $R$ from the beginning, but it does provide us a universal and convenient procedure for a large range of $p_T$'s. For these reasons, we are motivated to adopt the following procedure:
\begin{enumerate}
\item Use a large jet radius to cluster as many as $W$'s to single jets.
\item Apply the filtering/MD procedure to identify the two leading subjets. 
\item Collect jet constituents that are within two cones, each of which around one of the two subjet axes. The cone size is determined by the subjet's $p_T$,
\begin{equation}
R_{\text{sub}}= R_{\text{ref}}(100\gev)\frac{100\gev}{p_{T,\text{sub}}}
\end{equation}
where $R_{\text{ref}}(100\gev)$ is a reference radius at $p_T=100\gev$. In order to avoid excessively large cone sizes when one of the subjets has a small $p_T$, $R_{\text{sub}}$ is capped at 0.7. 
\item Use the jet constituents obtained in Step 3 to calculate jet radiation variables, and combine it with the (subtracted) filtered mass to get better discrimination.
\end{enumerate}
 
\begin{figure}[hbt!]
\centering
\includegraphics[width=0.6\textwidth]{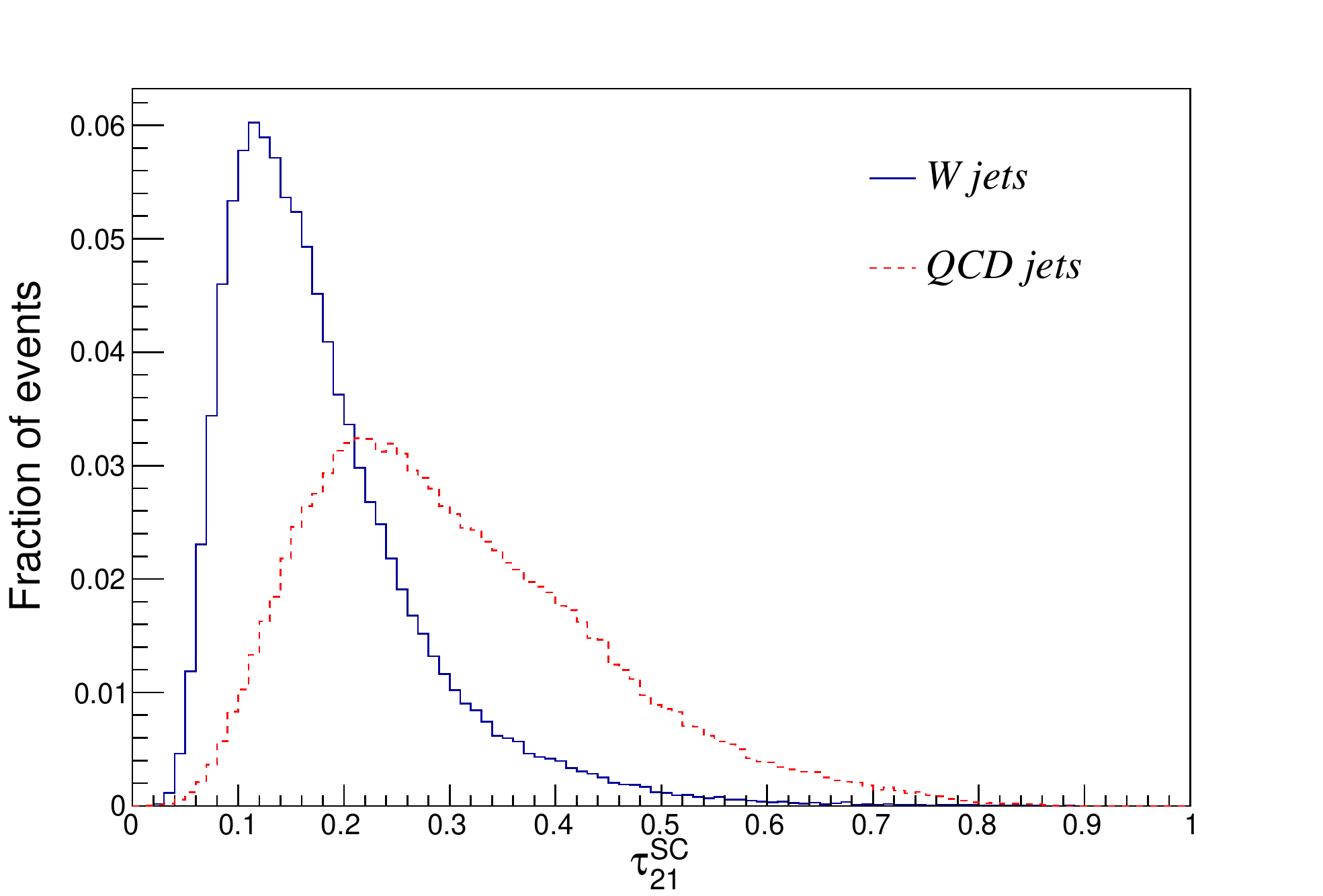}
\caption{The variable $\tau_{21}$ using the shrinking cone algorithm, $R_{\text{ref}}(100\gev) = 0.2$.\label{fig:cone}}
\end{figure}

Since the key ingredient in this procedure is in Step 3, we will call this method the ``shrinking cone'' (SC) algorithm. The $N$-subjettiness ratio, $\tau_2/\tau_1$, calculated using this procedure is denoted $\tau_{21}^\sc$. Choosing $R_{\text{ref}}(100\gev)=0.2$, we show the $\tau_{21}^\sc$ distributions for our $W$ jets and QCD jets in Fig.~\ref{fig:cone}. Similar to the previous subsection, we have applied a fixed mass window cut, $(60, 100)\gev$, on the subtracted filtered mass, and only included jets passing this cut in Fig.~\ref{fig:cone}. Comparing with the $\tau_{21}^\subtr$ distributions in Fig.~\ref{fig:subtraction}, we see a better distinction between $W$ jets and QCD jets. Applying a cut on $\tau_{21}^\sc$ such that $\es(\tau_{21}^\sc)=0.5$, we obtain $\eb(\tau_{21}^\sc)=0.10$ and $\es(\tau_{21}^\sc)/\sqrt{\eb(\tau_{21}^\sc)}=1.45$. As expected, we are unable to achieve the original performance of this variable in the zero pileup case. However, it does contribute to the discrimination power almost as much as the filtering/MD procedure. One may also wonder whether applying the pileup subtraction procedure will improve further the performance. We have tested it and found no improvement.

The signal and background efficiencies for various combinations of mass variables and $N$-subjettiness variables are shown in Table~\ref{tab:significance}.
\begin{table}[!htb]
  \centering
  \begin{tabular}{|c|c|c||c|c|c|}
  \hline
  \multicolumn{2}{|c|}{}&$\npu=0$&\multicolumn{3}{c|}{$\npu=60$}\\
  \hline\hline
  \multirow{4}{*}{Filtering/MD}&&$\mfilt\in (60,100)$&$\mfilt\in(80,120)$&$\mfilt^{\subtr}\in(60,100)$&$\mfilt^\subtr\in(60,100)$\\
\cline{2-6}
&$\es$& 0.64 & 0.48 & 0.60 & 0.60\\
\cline{2-6}
&$\eb$&0.12& 0.11 & 0.16 & 0.16\\
\cline{2-6}
&$\eseb$&1.84 & 1.47 & 1.51 & 1.51\\
\hline\hline
 \multirow{4}{*}{$\tau_{21}$}&&$\tau_{21}$&$\tau_{21}$&$\tau_{21}^{\subtr}$&$\tau_{21}^{\sc} (R_{\text{ref}} = 0.2)$\\
\cline{2-6}
&$\es$&0.50 (0.38)&0.50 (0.89) & 0.50 (0.66) & 0.50 (0.47)\\
\cline{2-6}
&$\eb$&0.10 (0.054)8& 0.25 (0.76) & 0.23 (0.37) & 0.12 (0.10)\\
\cline{2-6}
&$\eseb$&1.58 (1.61)& 1.00 (1.02)& 1.04 (1.08) & 1.45 (1.45)\\
\hline\hline
 \multirow{3}{*}{Total}&$\es$& 0.32 (0.24) & 0.24 (0.43)& 0.30 (0.39)& 0.30 (0.28)\\
\cline{2-6}
&$\eb$& 0.012 (0.0067) & 0.027 (0.082)& 0.037 (0.058)& 0.019 (0.016)\\
\cline{2-6}
&$\eseb$&2.92 (2.93) & 1.46 (1.49) & 1.57 (1.63) & 2.18 (2.18)\\
\hline\hline
\end{tabular}
\caption{Signal and background efficiencies and the improvement in $S/\sqrt{B}$. In the first step, ``filtering/MD'', we use fixed mass window cuts as shown in the second row. In the second step, we further cut on the $N$-subjettiness ratio for events within the mass window, and present the results for two choices of $\varepsilon_S$: 1. fixed $\varepsilon_S=0.5$; 2. (in the parentheses) $\varepsilon_S$ maximizing $\varepsilon_S/\sqrt{\varepsilon_B}$. The last group of numbers denoted ``Total'' are the products of the two steps, for example,  $\es(\text{total})= \es(m_\filt)\cdot\es(\tau_{21})$.\label{tab:significance}}
\end{table}

\begin{figure}[hbt!]
\centering
\includegraphics[width=0.7\textwidth]{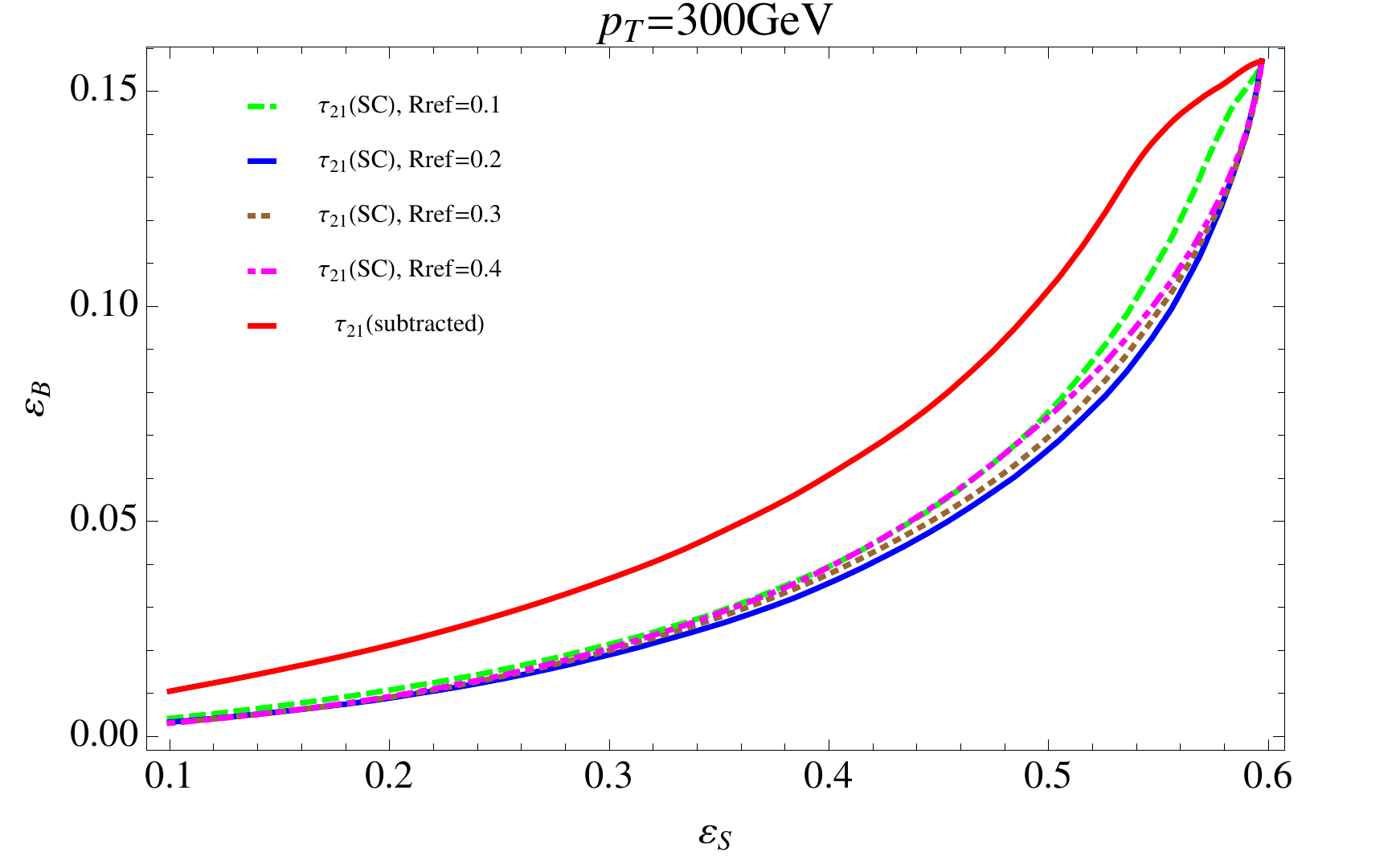}
\caption{Signal efficiency versus background fake rate for jet $p_T=300\gev$.\label{fig:roc_300}}
\end{figure}

A complete comparison between $\tau_{21}^\sc$ and $\tau_{21}^\subtr$ is given in Fig.~\ref{fig:roc_300}, where we plot the background fake rate as a function of the signal efficiency. When making the plots, we have again fixed the mass window cut as $60\gev<\mfilt^{\subtr}<100\gev$, which gives us $(\varepsilon_S,\varepsilon_B)=(0.60,0.16)$ as the maximum values at the top-right conner. Then we scan the cut on $\tau_{21}^\sc$ and $\tau_{21}^\subtr$ to produce the curves. It is seen that  $\tau_{21}^\sc$ is a better variable for all $\varepsilon_S$'s.

In Fig.~\ref{fig:roc_300}, we have also compared the performances of different choices of the reference radius: $\rref(100\gev)=0.1, 0.2, 0.3, 0.4$. It turns out for most of the signal efficiencies, $\rref(100\gev)=0.2$ is preferred. Nonetheless, as long as $\rref$ is not too large, the performance does not degrade significantly. This leaves room for practical cases where a very small radius is not viable, for example, when information from the hadronic calorimeter alone is used. It is also interesting to see if shrinking cones are better than cones of a fixed size. To see that, after obtaining the two subjets from the filtering/MD procedure, we use two cones of the same size to evaluate $\tau_{21}$, independent of the subjets' $p_T$. It turns out by using a cone size of 0.4 (0.7), we get $\es(\tau_{21})/\sqrt{\eb(\tau_{21})}=1.39$ (1.21) at $\es(\tau_{21})=0.5$. Comparing with the number from shrinking cones, 1.45, we see $R=0.4$ is almost as good, while the performance degrades significantly for $R=0.7$. If the pileup level is higher than that assumed in this article, the preferred jet radius will fall below those adopted by the LHC collaborations.

\begin{figure}[hbt!]
\centering
\begin{tabular}{cc}
\includegraphics[width=0.5\textwidth]{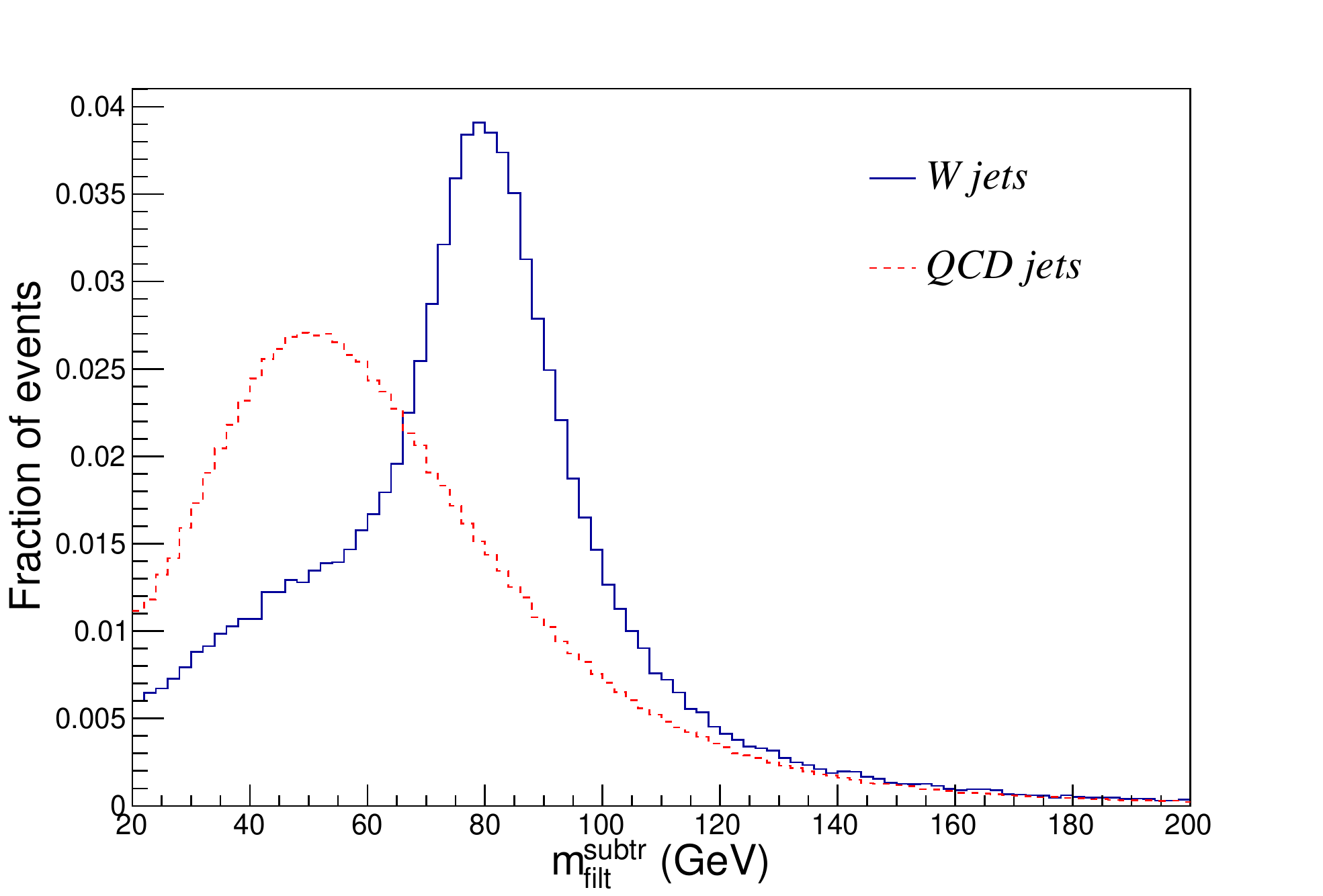}
&\includegraphics[width=0.5\textwidth]{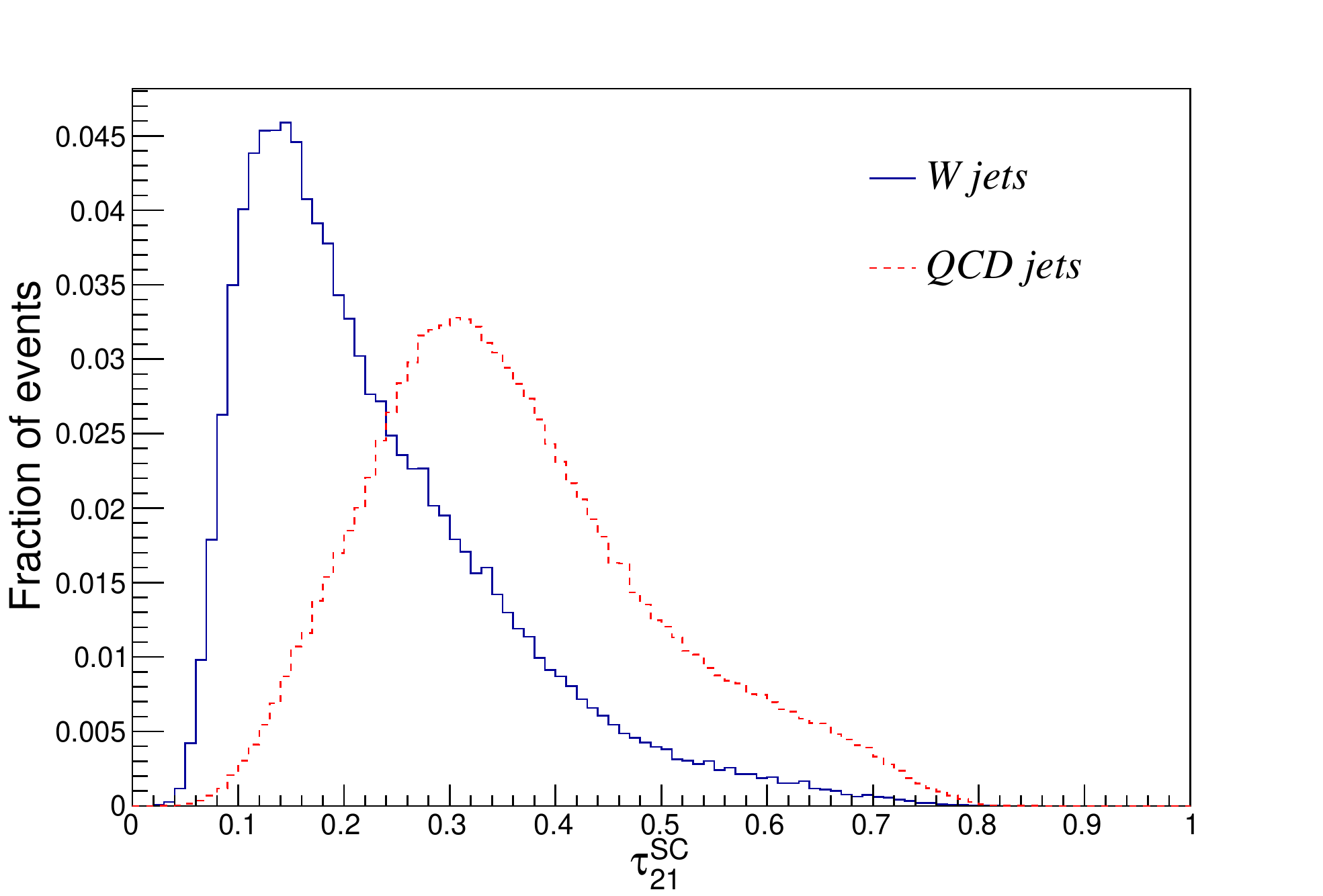}
\end{tabular}
\caption{The filtered mass after subtraction and $\tau_{21}$ using the the shrinking cone algorithm, for jet $p_T= 150\gev$. \label{fig:pt150}}
\end{figure}

In the above discussions, we have used $R=1.0$ to obtain the initial fat jet. If the $W$ $p_T$ is smaller, we will need a larger jet radius to cluster the $W$ decay products to a single jet\footnote{Alternatively, one may start with slim jets and find $W$'s by pairing jets with invariant masses close to the $W$ mass. Our method can be easily adapted accordingly.}. The contamination is even bigger because the jet area scales as $R^2$. In the following, we consider $W$'s with $p_T=150\gev$, clustered with $R=1.5$. The filtered mass after subtraction is given in the left panel of Fig.~\ref{fig:pt150}, for both $W$ jets and QCD jets. We see that, due to the larger jet radius, the filtering/MD procedure leaves a much larger portion of the background jets in the $W$ mass window: by choosing $60\gev<m_\filt^\subtr<100\gev$, we obtain $\varepsilon_S(m_\filt^\subtr)=0.54$ and $\varepsilon_B(m_\filt^\subtr)=0.31$, which gives us no increase in $S/\sqrt{B}$. Similarly, applying the pileup subtraction method on $\tau_{21}^\subtr$ does not provide any improvement either. On the other hand, the $\tau_{21}^{SC}$ variable are still efficient for separating the signal from the background, as shown in Fig~\ref{fig:pt150} (b). Choosing $R_\text{ref}(100\gev)=0.2$, we have $\varepsilon_B(\tau_{21}^\sc)=0.096$ at $\varepsilon_S(\tau_{21}^\sc)=0.5$, yielding $\es(\tau_{21}^\sc)/\sqrt{\eb(\tau_{21}^\sc)}=1.62$. This has made jet radiation patterns the most important handle for tagging semi-boosted $W$'s.

\begin{figure}[hbt!]
\centering
\includegraphics[width=0.7\textwidth]{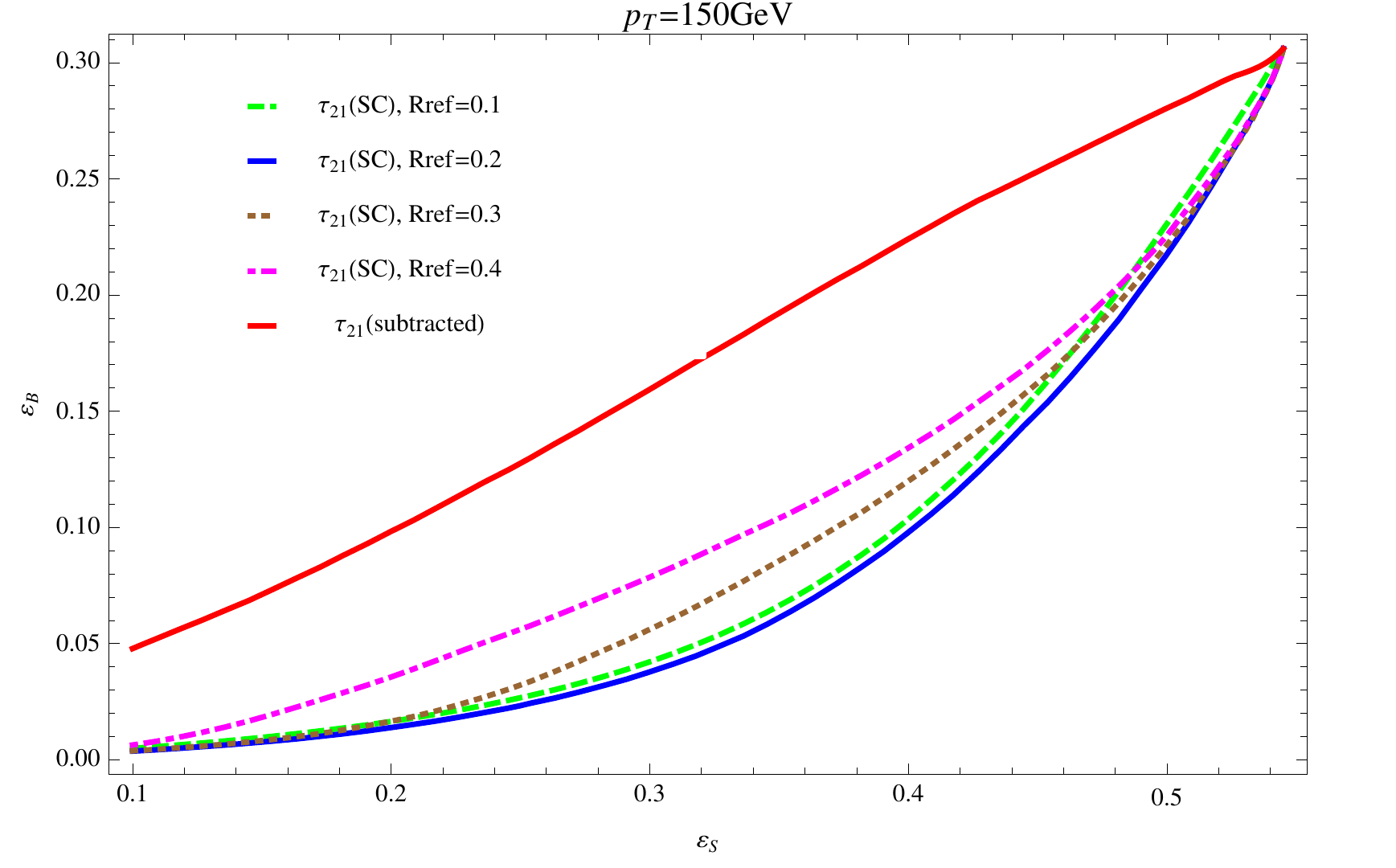}
\caption{Signal efficiency versus background fake rate for jet $p_T=150\gev$.\label{fig:roc_150}}
\end{figure}

 Similar to Fig.~\ref{fig:roc_300}, we plot the $\varepsilon_S\sim\varepsilon_B$ curves for several choices of $R_\text{ref}$ in Fig.\ref{fig:roc_150}, for jet $p_T=150\gev$. It turns out $R_\text{ref}(100\gev)=0.2$ is still the best choice among the values being considered. The difference between $p_T=150\gev$ and $p_T=300\gev$ is, the performance for $p_T=150\gev$ is getting worse faster when we increase $R_\text{ref}$. This is because for the same $R_\text{ref}$, the actual cone size is larger for lower $p_T$, and thus more contamination from pileup is included.
  
\section{Discussions}
\label{sec:discussions}
In this article, we have given a definition of the jet radiation radius, which quantifies the size of a jet due to its QCD radiation. This definition is closely related to the jet shape (also known as jet profile) variable which measures, on average, the fraction of momentum that is included in a cone of size $R$ around the jet axis. For the purpose of studying the jet radiation distribution, momentum is not a good measure because a large (small) momentum does not correspond to large (small) amount of radiation. Therefore, in our definition, we have replaced it with variables that directly measure the amount of radiation. Moreover, we have emphasized the nature of the radiation radius being {\it intrinsic}, {\it i.e.}, it is a characteristic of a parton with a definite QCD quantum, and should not change across different production processes and experimental setups. In particular, it is defined before various contaminations are included. It is our hope that by ``factorizing'' the contributions to jet shape variables into intrinsic and environmental ones, we can simplify jet substructure studies and use jet radiation patterns more efficiently to distinguish jets with different quantum numbers.

A key observation in this article is: in order to efficiently use jet radiation variables, a smaller than usual jet radius is often preferred. This is particularly true for the $W$ jet tagging method we proposed, where shrinking cones are used when calculating jet radiation variables. In a high energy experiment, either a simpler or a more complicated method may be adopted. On the one hand, due to the limitation in granularities, especially those of the hadronic calorimeter, one may not be able to use a jet radius smaller than $O(0.1)$, or/and may not be able to use continuous jet radii in the calculation. In that case, we may choose to simplify the method by choosing a few typical, but small, cone sizes. It is shown in Fig.~\ref{fig:roc_300} and Fig.~\ref{fig:roc_150} that increasing the jet radius within a sizable region will not significantly hurt the discrimination power. On the other hand, one does see the advantages of using small cones, for example, $R_\text{sub}=0.2$ is preferred for subjet $p_T\sim100\gev$ when the average number of pileup events is 60. An even smaller radius might be preferred if the pileup level is higher. Therefore, ideally we would want to use the finest granularity for jet constituents, including the information from the electromagnetic calorimeter and the tracking system as in the particle flow approach \cite{pf}. Moreover, we have sticked to a single choice of (sub)jet radius for each (sub)jet in this article. Similar to Ref.~\cite{wtag}, one may benefit from using two or more radii for each (sub)jet, which not only gives us the information of how much the radiation is, but also captures how it grows with increasing $R$.

The shrinking cone algorithm we proposed for $W$ tagging is parallel and complementary to other pileup reduction methods and may be combined to obtain the optimum results. We have already used it with the pileup subtraction method proposed in Ref.~\cite{subtraction}, where we see the subtraction method is convenient to extract the kinematic information while the shrinking cone method is more useful to obtain the radiation information. Another set of useful techniques utilize the fact that a charged particle from a pileup event leaves a track not originated from the primary vertex, thus it can be subtracted from the jet. These methods include charged hadron subtraction \cite{chs}, using a jet vertex fraction \cite{jvf} cut and jet cleansing \cite{cleansing}. One may even use the charged particles from the primary vertex alone when calculation radiation variables, which still provides us a lot of information for the color structure \cite{tracking}. To improve over these method, one may simply apply them for cones with sizes determined by the (sub)jet $p_T$'s, as discussed in this article. Here, we emphasize that even if we only use tracks from the primary vertex to avoid most of the contamination from pileup events, it is still useful to optimize the jet cone sizes. We have seen that it is the case for quark-gluon discrimination when pileup is turned off. We expect this consideration to be more important when dealing with events with many hard partons, where jets can be easily contaminated by nearby radiation and a large jet radius should be avoided. This happens in, for example, SUSY cascades with long decay chains. 

In conclusion, we have shown that the knowledge of the intrinsic jet radiation radius will lead us towards the optimum discriminations for jets with different quantum numbers.

\acknowledgments
The author thanks Dave Soper for numerous useful discussions and comments on the manuscript. The author is in part supported by US Department of Energy under grant numbers DE-FG02-96ER40969 and DE-FG02-13ER41986. 

\end{document}